\documentclass[a4paper,11pt]{article}
\usepackage{jheppub} 

\usepackage{amsmath, amssymb}
\usepackage{graphicx}
\usepackage{hyperref}
\usepackage{enumerate}
\usepackage{longtable}
\usepackage{titlesec}
\usepackage{empheq}
\usepackage[most]{tcolorbox}

\title{2D CFT and efficient Bethe ansatz for exactly solvable Richardson--Gaudin models}

\author[]{Grzegorz Biskowski,}
\author[]{Franco Ferrari,}
\author[]{Marcin R.~Pi\c{a}tek}

\affiliation[]{Institute of Physics, University of Szczecin,  Wielkopolska 15, 70--451 Szczecin, Poland}

\emailAdd{biskowski.grzegorz@gmail.com}
\emailAdd{franco.ferrari@usz.edu.pl}
\emailAdd{marcin.piatek@usz.edu.pl}

\abstract{
This work inaugurates a series of complementary studies on Richardson--Gaudin integrable models. We begin by reviewing the 
foundations of classical and quantum integrability, recalling the algebraic Bethe ansatz solution of the Richardson (reduced 
BCS) and Gaudin (central spin) models, and presenting a proof of their integrability based on the Knizhnik--Zamolodchikov 
equations and their generalizations to perturbed affine conformal blocks. Building on this 
foundation, we then describe an alternative CFT-based formulation. In this approach, the Bethe ansatz equations for these exactly 
solvable models are embedded within two-dimensional Virasoro CFT via irregular, degenerate conformal blocks.
To probe new formulations within the Richardson--Gaudin class, we develop a high-performance numerical solver. 
The Bethe roots are encoded in the Baxter polynomial, with initial estimates obtained from a secular matrix 
eigenproblem and subsequently refined using a deflation-assisted hybrid Newton--Raphson/Laguerre algorithm.
The solver proves effective in practical applications: when applied to picket-fence, harmonic oscillator, and hydrogen-like 
spectra, it accurately reproduces known rapidity trajectories and 
reveals consistent merging and branching patterns of arcs in the complex rapidity plane. 
We also explain how to generalize our computational approach to finite temperatures, allowing 
us to calculate temperature-dependent pairing energies and other thermodynamic observables directly within the discrete 
Richardson model. We propose an application of the solver to Gaudin-type Bethe equations, which emerge in the classical 
(large central charge) limit of Virasoro conformal blocks. We conclude by outlining future directions: direct minimization of the 
Yang--Yang function as an alternative root-finding strategy; revisiting time-dependent extensions; and 
exploration of complementary analytic frameworks, including matrix models, 2D CFT techniques, and 4D gauge theory/2D CFT 
dualities.
}

\keywords{Conformal and W Symmetry, Bethe Ansatz}

\allowdisplaybreaks
\newcommand{\ket}[1]{\left|\, #1 \,\right\rangle}

\newcommand{\To}{\longrightarrow}

\begin{document}
\maketitle


\newpage

\section{Introduction}
Quantum integrable models occupy a central place in the study of strongly correlated many-body systems, offering exact analytical 
control over non-perturbative phenomena ranging from superconductivity to nuclear pairing and quantum magnetism.  Among these, 
the Richardson--Gaudin (RG) family of models \cite{Dukelsky:2004re,Claeys:2018zwo} 
provides a paradigmatic example: the exactly solvable 
pairing Hamiltonian introduced by Richardson and Sherman \cite{Richardson1963, Richardson1964} describes fermions interacting 
through an attractive pairing force, while the Gaudin spin magnets \cite{Gaudin1976, Gaudin2014} capture the dynamics of spins 
coupled via long-range exchange. These models admit Bethe ansatz solutions, in which the many-body spectrum is encoded in a set 
of nonlinearly coupled algebraic equations for rapidities (Bethe roots).

The primary objective of this paper is to review the standard Bethe ansatz formulation of Richardson--Gaudin models and to 
reexamine their established connection with two-dimensional conformal field theory (2D CFT) as described in the seminal work 
\cite{Sierra:1999mp}. 
Building on this foundation, we present a CFT-based realization of the Yang--Yang (YY) 
function that offers an alternative to the construction proposed in \cite{Sierra:1999mp}
by employing irregular, degenerate Virasoro conformal 
blocks.\footnote{Recall that the YY function 
serves as the generating potential for the Bethe equations.} 
This representation offers new analytic tools for studying the spectra of RG models.\footnote{To our knowledge, 
this observation has not been explicitly addressed in the context of Richardson--Gaudin models.}

Beyond their intrinsic mathematical elegance, RG models find wide application across condensed matter and nuclear 
physics. The reduced BCS Hamiltonian lies at the heart of superconductivity in metallic grains and governs the crossover from 
few-to many-particle pair excitations in ultracold atomic gases confined to small reservoirs 
\cite{VonDelft1999,OrtizDukelsky,ResareHofmann}---situations in which discrete energy levels demand an exact treatment.
Gaudin magnets, in turn, provide insight into spin wave dynamics in quantum dots and central spin decoherence in spin-qubit 
devices \cite{He2022}. In all cases, extracting physical observables such as ground state energy, excitation spectra, and 
correlation functions relies on the numerical solution of Richardson (Bethe) equations as a function of the coupling constant 
\(g\).

The second objective of this work is to develop robust numerical methods for tracking rapidities as the coupling $g$ varies, with 
direct applications to the Richardson equations.  This paper also serves as a report on the implementation and testing of these 
methods and includes a link to the publicly available code (see subsection~\ref{code}), which we release for community use. 
Our numerical procedure comprises two steps: (i) an eigenvalue-based 
method that reformulates the Bethe equations as the logarithmic derivative of the Baxter polynomial \cite{Faribault2011}; and 
(ii) determining the coefficients of the Baxter polynomial using the Newton--Raphson method combined with LU decomposition, 
followed by direct root finding via the Laguerre method, enhanced through systematic polynomial deflation.

Our goal was to develop a flexible code that can be easily adapted to different tasks. For example, it reproduces known 
numerical dependencies of the rapidities on $g$ in both non-degenerate and degenerate regimes. It also supports testing new 
hypotheses---such as the correspondence between RG models and 2D CFT or other novel formulations described in the conclusions.

The paper is organized as follows. In section~\ref{Ch2}, with an emphasis on self-contained exposition, we review the foundations 
of classical and quantum integrability and introduce the algebraic Bethe ansatz. This section draws on the comprehensive review 
\cite{Retore2022} and references therein. 
In section~\ref{Ch3}, we present the Richardson--Gaudin models: we derive the Richardson equations, discuss the connection 
between the reduced BCS model and Gaudin spin systems (informed by 
\cite{Dukelsky:2004re,Claeys:2018zwo,VonDelft1999,Cambiaggio:1997vz,Sierra:1999mp,Sierra2001}), and describe the embedding of 
these models within 2D CFT \cite{Sierra:1999mp}. We also introduce our key observation: an alternative representation of the 
Yang--Yang function using irregular Virasoro blocks.
In section~\ref{Ch4}, we review the eigenvalue-based method for solving the Richardson equations via their correspondence to 
ordinary differential equations, following \cite{Faribault2011}. 
Section~\ref{Ch5} details our numerical implementation---based on techniques from \cite{NumericalRecipes}---and describes the 
code architecture. Section~\ref{Ch6} presents our numerical results, including graphical visualizations of ground-state 
rapidities as functions of the coupling \(g\) across various models and parameter regimes. We then propose an extension of our 
computational framework to finite temperatures, enabling direct computation of temperature-dependent pairing energies and other 
thermodynamic observables in the discrete Richardson model, and outline how the solver can be applied to Gaudin-type Bethe 
equations arising in the classical limit of Virasoro conformal blocks.
Finally, section~\ref{Ch7} summarizes our main findings and outlines prospects for future research.

\section{Integrability and the algebraic Bethe ansatz}
\label{Ch2}
\subsection{Classical integrability}
Let us consider a \( 2n \)-dimensional phase space with canonical coordinates \( \{q_i\} \) and conjugate momenta \( \{p_i\} \), 
for \( i = 1, \ldots, n \). For two sufficiently smooth functions \( F(q_i, p_i) \) and \( G(q_i, p_i) \) defined on this phase 
space, the {\it Poisson bracket} \( \{F, G\} \) is defined as
\begin{equation}
\{F, G\} \equiv \sum_{i=1}^{n} \left( \frac{\partial F}{\partial q_i} \frac{\partial G}{\partial p_i} - \frac{\partial F}
{\partial p_i} \frac{\partial G}{\partial q_i} \right).
\end{equation}
The Poisson bracket satisfies several important properties: it is bilinear, antisymmetric,
$$
\{F, G\} = -\{G, F\},
$$
and obeys the Jacobi identity,
$$
\{F, \{G, H\}\} + \{G, \{H, F\}\} + \{H, \{F, G\}\} = 0,
$$
for any smooth functions \( F, G, H \). 
These properties make the Poisson bracket a fundamental structure in the study of 
classical integrable systems.

Let us consider a smooth function \( F(\{q_i\}, \{p_i\}) \) defined on a \( 2n \)-dimensional phase space. 
Its time derivative is given by
\begin{align}\label{eq: 2_clas_int-Fdot}
\dot{F}\equiv\frac{{\rm d}F}{{\rm d}t}= 
\frac{{\rm d}}{{\rm d}t}\left[\sum_{i=1}^{n}\left(\frac{\partial F}{\partial q_i}\,{\rm d}q_i 
+\frac{\partial F}{\partial p_i}\,{\rm d}p_i\right)\right] 
= \sum_{i=1}^{n}\left(\frac{\partial F}{\partial q_i} \frac{{\rm d}q_i}{{\rm d}t} 
+\frac{\partial F}{\partial p_i} \frac{{\rm d}p_i}{{\rm d}t}\right).
\end{align}
According to Hamilton's canonical equations, we have
\begin{align}
\frac{{\rm d}q_i}{{\rm d}t} = \frac{\partial H}{\partial p_i}, \qquad 
\frac{{\rm d}p_i}{{\rm d}t} = -\frac{\partial H}{\partial q_i}.
\end{align}
Substituting these into equation~\eqref{eq: 2_clas_int-Fdot}, we obtain
\begin{align}\label{eq: 2_clas_int-FH_bracket}
\dot{F} = \sum_{i=1}^{n} \left( \frac{\partial F}{\partial q_i} \frac{\partial H}{\partial p_i} - \frac{\partial F}{\partial p_i} 
\frac{\partial H}{\partial q_i} \right) 
=\{F, H\},
\end{align}
where \( \{F, H\} \) denotes the Poisson bracket of \( F \) and \( H \).

If the Poisson bracket of two functions vanishes, i.e., \( \{F, G\} = 0 \), we say that the functions \( F \) and \( G \) are 
{\it in involution}. From equation~\eqref{eq: 2_clas_int-FH_bracket}, it follows that if a function \( F \) is in involution 
with the Hamiltonian \( H \), then its time derivative vanishes, \( \dot{F} = 0 \). Consequently, \( F \) is a conserved 
quantity, i.e., it remains constant along the dynamical trajectories.

A system defined on a \( 2n \)-dimensional phase space is said to be {\it Liouville integrable} if there exist \( n \) 
functionally independent conserved quantities \( \{F_1, \ldots, F_n\} \), including the Hamiltonian \( H \), which are all 
in mutual involution:
\begin{equation}
\{F_i, F_j\} = 0 \qquad \forall\; i,j = 1, \ldots, n.
\end{equation}

According to {\it Liouville's theorem}, the equations of motion of such a system can be solved by quadratures.\footnote{
One typically defines \textit{quadratures} as solutions obtained by a finite sequence of algebraic operations and integrations.}
In particular, there exists a canonical transformation to so-called \textit{action--angle variables}, 
in which the dynamics becomes particularly simple and transparent.\footnote{
The terminology \textit{action-angle variables} refers to the fact that the new variables consist of conserved quantities 
(actions) and their canonically conjugate cyclic variables (angles). These variables appear naturally in integrable systems.}
This transformation takes the form
\begin{equation}
\left(q_i, p_i\right)\To\left(F_i, \varphi_i\right),
\end{equation}
where \( \{F_i\} \) is the set of conserved action variables, and \( \{\varphi_i\} \) are their conjugate angle variables.
It then follows from Hamilton's equations that
\begin{align}
\dot{F}_i &= \{F_i, H\} = 0, \\
\dot{\varphi}_i &= \{\varphi_i, H\} \equiv \Omega_i,
\end{align}
where \( \Omega_i \) are constant frequencies determined by the Hamiltonian.
The general solution of the system in action-angle coordinates is then
\begin{align}
F_i(t) &= \alpha_i = \text{const.}, \\
\varphi_i(t) &= \Omega_i t + \varphi_i(0),
\end{align}
showing that the motion is linear and uniform in the angle variables, while the actions remain constant in time.

Let us now consider two matrices \(L\) and \(M\), referred to as the \textit{Lax pair}, satisfying the Lax equation
\begin{equation}\label{eq:2_clas_int-Ldot_comm}
\dot{L} = [M, L],
\end{equation}
where \( [\cdot,\cdot] \) denotes the matrix commutator. 

It can be shown that this formulation implies the existence of conserved quantities \( Q_n \), defined as
\begin{equation}\label{Qn}
Q_n = \operatorname{tr}(L^n),
\end{equation}
which remain constant in time. 
These quantities are in involution and play a fundamental role in the integrability of the system.
Let us prove that the quantities (\ref{Qn}) are conserved under the Lax equation (\ref{eq:2_clas_int-Ldot_comm}).
We compute the time derivative of \( Q_n \):
\begin{align}
\frac{{\rm d}}{{\rm d}t} Q_n &=\frac{{\rm d}}{{\rm d}t} \operatorname{tr}(L^n) 
=\operatorname{tr}\left( \frac{{\rm d}}{{\rm d}t} L^n \right).
\end{align}
Using the identity for the derivative of a matrix power:
\[
\frac{{\rm d}}{{\rm d}t} L^n = \sum_{k=0}^{n-1} L^k \dot{L} L^{n-1-k},
\]
one gets
\begin{align}
\frac{{\rm d}}{{\rm d}t} Q_n &= \operatorname{tr} \left( \sum_{k=0}^{n-1} L^k \dot{L} L^{n-1-k} \right) 
= \sum_{k=0}^{n-1} \operatorname{tr} \left( L^k [M, L] L^{n-1-k} \right).
\end{align}
Now let us use the cyclic property of the trace, \( \operatorname{tr}(AB) = \operatorname{tr}(BA) \), and the fact that 
\(\operatorname{tr}([A, B]) = 0\) for any matrices \( A, B \). Each term in the sum is a trace of a commutator:
\begin{equation}
\operatorname{tr}(L^k [M, L] L^{n-1-k}) 
=\operatorname{tr}\left( [M, L] L^{n-1} \right) = \operatorname{tr}([M, L^n]) = 0.
\end{equation}
Therefore,
\[
\frac{{\rm d}}{{\rm d}t} Q_n = 0,
\]
which proves that \( Q_n = \operatorname{tr}(L^n) \) is conserved.

We can verify that equation~(\ref{eq:2_clas_int-Ldot_comm}) admits solutions of the form
\begin{equation}\label{eq: 2_clas_int-pair_sols}
L(t) = g(t) L(0) g(t)^{-1}, \qquad M(t) = \frac{{\rm d}g(t)}{{\rm d}t} g(t)^{-1},
\end{equation}
where \(g(t)\) is an invertible matrix-valued function of time.
To confirm this, let us differentiate \(L(t)\) with respect to time:
\begin{align}
\dot{L}(t) &= \frac{{\rm d}}{{\rm d}t}\left[g(t) L(0) g(t)^{-1}\right] \nonumber \\
           &= \dot{g}(t) L(0) g(t)^{-1} + g(t) L(0) \frac{{\rm d}}{{\rm d}t}\left(g(t)^{-1}\right).
\end{align}
We now use the identity
\[
\frac{{\rm d}}{{\rm d}t}g(t)^{-1} = -g(t)^{-1} \dot{g}(t) g(t)^{-1},
\]
to obtain
\begin{align}
    \dot{L}(t) &= \dot{g}(t) L(0) g(t)^{-1} - g(t) L(0) g(t)^{-1} \dot{g}(t) g(t)^{-1} \nonumber \\
               &= \dot{g}(t) g(t)^{-1} \cdot L(t) - L(t) \cdot \dot{g}(t) g(t)^{-1} \nonumber \\
               &= \left[\dot{g}(t) g(t)^{-1},\, L(t)\right].
\end{align}
Finally, identifying \( M(t) = \dot{g}(t) g(t)^{-1} \), we obtain
\begin{equation}\label{LaxEofM}
  \dot{L}(t) = [M(t), L(t)],
\end{equation}
which confirms that the ansatz (\ref{eq: 2_clas_int-pair_sols}) satisfies the Lax equation~(\ref{eq:2_clas_int-Ldot_comm}).
This implies that the eigenvalues of the Lax matrix 
$L$ are preserved under time evolution. Consequently, equation~(\ref{eq:2_clas_int-Ldot_comm}) is classified as an {\it 
isospectral flow}, reflecting the fact that the spectrum of 
$L$ is invariant with respect to time.

Indeed, the term isospectral flow refers to a time evolution of a matrix (or more generally, a linear operator) such 
that its spectrum --- the set of its eigenvalues --- remains invariant in time.
This concept naturally arises in the theory of integrable systems through the Lax pair formulation. A dynamical system 
admits a Lax representation if there exist matrices \( L(t) \) and \( M(t) \) such that the equations of motion can be written in 
the form (\ref{LaxEofM}). 
The structure of this equation ensures that \( L(t) \) evolves by a similarity transformation:
\begin{equation}
    L(t) = g(t) L(0) g(t)^{-1},
\end{equation}
for some invertible matrix \( g(t) \). As similarity transformations preserve the spectrum of a matrix, it follows that all 
eigenvalues of \( L(t) \) are conserved quantities. 
Isospectral flows play a fundamental role in integrable systems, as the conserved quantities --- which guarantee integrability in 
the sense of Liouville --- are often encoded in the eigenvalues of the Lax matrix \( L \).\footnote{In the Lax pair formulation 
of integrable systems, the conserved quantities are always encoded in the eigenvalues of the Lax matrix \( L \), since the 
isospectral condition \( \dot{L} = [M, L] \) ensures that these eigenvalues remain constant in time. Thus, functions of the 
spectrum, such as \( \operatorname{tr} L^n \), yield conserved quantities. However, in more general settings, additional or 
alternative conserved quantities may arise from {\it monodromy} or {\it transfer matrices}, 
and not all integrable systems necessarily admit a Lax representation with this property.}

Let us define the \textit{Kronecker product} of two \( n\times n \) matrices \(A\) and \(B\). 
The Kronecker product \(A\otimes B\) is given by
\begin{equation}
    A\otimes B = \begin{pmatrix}
        a_{11}B & a_{12}B & \cdots & a_{1n}B \\
        a_{21}B & a_{22}B & \cdots & a_{2n}B \\
        \vdots & \vdots & \ddots & \vdots \\
        a_{n1}B & a_{n2}B & \cdots & a_{nn}B
    \end{pmatrix} = \begin{pmatrix}
        a_{11}b_{11} & a_{11}b_{12} & \cdots & a_{1n}b_{1n}\\
        a_{11}b_{21} & a_{11}b_{22} & \cdots & a_{1n}b_{2n}\\
        \vdots & \vdots & \ddots & \vdots \\
        a_{n1}b_{n1} & a_{n1}b_{n2} & \cdots & a_{nn}b_{nn}
    \end{pmatrix}.
\end{equation}
This operation produces a block matrix where each entry \( a_{ij} \) of \( A \) is multiplied by the entire matrix \( B \).

Let us now consider the tensor product space \( V^{\otimes N} = V \otimes V \otimes \cdots \otimes V \), consisting of \( N \) 
copies of the vector space \( V \). We introduce the notation \( A_i \) to indicate that the operator \( A \) acts non-trivially 
only on the \( i \)-th site and as the identity elsewhere. That is,
\begin{equation} \label{eq: 2_clas_int-site_op}
 A_i = \mathbb{I}^{\otimes (i-1)} \otimes A \otimes \mathbb{I}^{\otimes (N-i)},
\end{equation}
where \( \mathbb{I} \) is the identity operator on \( V \).

From this definition, it is straightforward to verify that two operators acting on different sites commute:
\begin{equation}
    i \neq j \quad \Rightarrow \quad [A_i, A_j] = 0.
\end{equation}

This construction allows us to naturally define a matrix \( L_1 \) in the extended space as
\begin{equation}
 L_1 = L \otimes \mathbb{I},
\end{equation}
where \( L \) acts on the first copy of \( V \), and \( \mathbb{I} \) acts on an auxiliary or second space. 

Let us now assume that the Lax matrix \( L \) can be diagonalized by a similarity transformation using some invertible matrix \( U \):
\begin{equation}
    L = U \Lambda U^{-1},
\end{equation}
where \( \Lambda \) is a diagonal matrix. This implies that the eigenvalues \( \Lambda_{ii} \) are conserved quantities, and hence
\begin{equation}
    \left\{\Lambda_i, \Lambda_j\right\} = 0, \qquad \forall\, i, j,
\end{equation}
where the indices \( i \) and \( j \) label the sites, consistent with the notation introduced in equation~(\ref{eq: 2_clas_int-site_op}).

Using this property, we can express the Poisson bracket \( \left\{L_1, L_2\right\} \) by substituting the decomposed forms:
\begin{equation}
    L_1 = U_1 \Lambda_1 U_1^{-1}, \qquad L_2 = U_2 \Lambda_2 U_2^{-1}.
\end{equation}
As shown in appendix \ref{A}, this leads to the following fundamental Poisson bracket relation:
\begin{equation}
    \left\{L_1, L_2\right\} = \left[r_{12}, L_1\right] - \left[r_{21}, L_2\right],
\end{equation}
where \( r_{12} \) is known as the classical \( r \)--matrix.

We can now insert this result into the Jacobi identity involving a triple of Lax matrices \( \{L_1, L_2, L_3\} \), which yields:
\begin{align}\label{eq: 2_clas_int-jaco-r}
    &\left[L_1, \left\{L_2, r_{13}\right\} - \left\{L_3, r_{12}\right\} + \left[r_{12}, r_{13} + r_{23}\right] + \left[r_{32}, r_{13}\right] \right] + \nonumber\\
    &\left[L_2, \left\{L_3, r_{21}\right\} - \left\{L_1, r_{23}\right\} + \left[r_{23}, r_{21} + r_{31}\right] + \left[r_{13}, r_{21}\right] \right] + \nonumber\\
    &\left[L_3, \left\{L_1, r_{32}\right\} - \left\{L_2, r_{31}\right\} + \left[r_{31}, r_{32} + r_{12}\right] + \left[r_{21}, r_{32}\right] \right] = 0.
\end{align}

In the special case where the \( r \)--matrices are independent of the dynamical variables and antisymmetric, i.e., \( r_{ij} = -r_{ji} \), the above expression simplifies and leads to the celebrated \textit{classical Yang--Baxter equation} (CYBE):
\begin{equation}
    \left[r_{12}, r_{13} + r_{23}\right] + \left[r_{32}, r_{13}\right] = 0.
\end{equation}
This structure is central to the classical \(r\)--matrix formalism and underlies many integrable systems.

As a conclusion to our discussion of classical integrability, we summarize the key features of the classical 
$r$--matrix formalism.

\paragraph{Summary of the classical $r$--matrix formalism.}
The classical \(r\)--matrix formalism provides a powerful algebraic framework for studying integrable systems in the Hamiltonian 
formulation. It encodes the Poisson bracket relations between the elements of the Lax matrix and ensures the integrability of the 
model via the existence of conserved quantities in involution.

Let \( L \) be the Lax matrix of a classical integrable system. To describe its Poisson algebraic structure, we introduce the 
tensor product notation:
\begin{equation}
L_1 = L \otimes \mathbb{I}, \qquad L_2 = \mathbb{I} \otimes L,
\end{equation}

The Poisson brackets of the matrix elements of \(L\) can be compactly written using an object called the classical 
\( r \)--matrix, denoted by \( r_{12}(u,v) \in \text{End}(V \otimes V) \), satisfying
\begin{equation} \label{eq: r_bracket}
\{L_1(u), L_2(v)\} = [r_{12}(u,v), L_1(u)] - [r_{21}(u,v), L_2(v)],
\end{equation}
where \(u, v\) are spectral parameters and \(r_{21}(u,v)\equiv P r_{12}(v,u) P \), with \(P\) being the permutation operator on 
\(V\otimes V\).

\medskip\noindent
{\bf Isospectrality and integrals of motion.} From the structure of \eqref{eq: r_bracket}, it follows that the traces \( \text{tr}(L^n(u)) \) form a commuting family of conserved quantities:
\begin{equation}
    \{\text{tr}(L^n(u)), \text{tr}(L^m(v))\} = 0 \qquad \forall\; m,n.
\end{equation}
Thus, the existence of an \( r \)--matrix satisfying \eqref{eq: r_bracket} guarantees Liouville integrability of the system.

\medskip\noindent
{\bf Classical Yang--Baxter equation.} 
The consistency of this Poisson structure is ensured by the classical Yang--Baxter equation:
\begin{equation}
    [r_{12}, r_{13}] + [r_{12}, r_{23}] + [r_{13}, r_{23}] = 0 \in \text{End}(V^{\otimes 3}),
\end{equation}
which the \( r \)--matrix must satisfy.

\subsection{Quantum integrability}
For quantum integrability, we assume the existence of a family of mutually commuting conserved charges \(Q_i\),  
\begin{equation}
[Q_i, Q_j] \;=\; 0\,,\qquad i,j=1,2,\ldots\;.
\end{equation}
Our task is to construct these charges from the underlying integrable structure.

To this end, we introduce a quantum \(R\)\nobreakdash-matrix, an operator  
\begin{equation}
    R(z_1,z_2)\;\colon\;\mathrm{End}\bigl(\mathcal{H}\otimes\mathcal{H}\bigr)\;\longrightarrow\;\mathrm{End}\bigl(\mathcal{H}\otimes\mathcal{H}\bigr),
\end{equation} 
which satisfies the {\it quantum Yang--Baxter equation} (QYBE):
\begin{equation}
    R_{12}(z_1,z_2)\,R_{13}(z_1,z_3)\,R_{23}(z_2,z_3)
    \;=\;
    R_{23}(z_2,z_3)\,R_{13}(z_1,z_3)\,R_{12}(z_1,z_2)\,,
\end{equation}
with spectral parameters \(z_i\in\mathbb{C}\).  We will restrict to the additive case  
\begin{equation}
    R_{ij}(z_i,z_j)\;=\;R_{ij}(z_i - z_j)\,.
\end{equation}
Shifting \(z_1\To z_1+z_3\) and \(z_2\To z_2+z_3\) then yields the more familiar form
\begin{equation}
    R_{12}(z_1 - z_2)\,R_{13}(z_1)\,R_{23}(z_2)
    \;=\;
    R_{23}(z_2)\,R_{13}(z_1)\,R_{12}(z_1 - z_2)\,.
\end{equation}

Next, we introduce a Lax operator \(\mathcal{L}_i\) that relates the state at site \(i\) to the state at site \(i+1\):
\begin{equation}
\ket{v_{i+1}} = \mathcal{L}_i\ket{v_i}.
\end{equation}
Equivalently, \(\mathcal{L}_i\) satisfies the RLL relation
\begin{equation}
    R_{\alpha\beta}(z_\alpha - z_\beta)\,\mathcal{L}_{\alpha j}(z_\alpha)\,\mathcal{L}_{\beta j}(z_\beta)
    \;=\;
    \mathcal{L}_{\beta j}(z_\beta)\,\mathcal{L}_{\alpha j}(z_\alpha)\,R_{\alpha\beta}(z_\alpha - z_\beta),
\end{equation}
where \(\mathcal{L}_{\alpha j}(z)\) acts on the tensor product \(\mathcal{V}_\alpha\otimes\mathcal{H}_j\), with \(\mathcal{V}_\alpha\) the auxiliary space (labeled by \(\alpha\)) and \(\mathcal{H}_j\) the physical Hilbert space at site \(j\).  

For a fixed \(R\)\nobreakdash-matrix, the RLL equation generally admits multiple solutions for the Lax operator \(\mathcal{L}(z)\).  A canonical solution is simply
\begin{equation}
    \mathcal{L}(z) \;=\; R(z)\,,
\end{equation}
which we adopt for concreteness in what follows.

We next assemble the Lax operators into the {\it monodromy matrix}, which transports states from site \(1\) to site \(N\):
\begin{equation}
    \mathcal{T}_{\alpha}(z) \;=\; \mathcal{L}_{\alpha N}(z)\,\mathcal{L}_{\alpha\,N-1}(z)\,\cdots\,\mathcal{L}_{\alpha 2}(z)\,\mathcal{L}_{\alpha 1}(z).
\end{equation}
One shows, using repeated applications of the RLL relation, that \(\mathcal{T}\) satisfies
\begin{equation}
    R_{\alpha\beta}(z_\alpha - z_\beta)\,\mathcal{T}_{\alpha}(z_\alpha)\,\mathcal{T}_{\beta}(z_\beta)
    = 
    \mathcal{T}_{\beta}(z_\beta)\,\mathcal{T}_{\alpha}(z_\alpha)\,R_{\alpha\beta}(z_\alpha - z_\beta).
\end{equation}
Taking the trace over the auxiliary space \(\mathcal{V}_\alpha\) defines the {\it transfer matrix}
\begin{equation}
    t(z) \;=\; \operatorname{tr}_{\alpha}\mathcal{T}_{\alpha}(z),
\end{equation}
which generates the mutually commuting conserved charges.  Indeed, cyclicity of the trace implies
\begin{equation}
    \bigl[t(z_1),\,t(z_2)\bigr] \;=\; 0 \quad\forall\,z_1,z_2.
\end{equation}
Expanding its logarithm in powers of \(z\),
\begin{equation}
    \log t(z) \;=\;\sum_{\ell=0}^\infty Q_\ell\,z^\ell,
\end{equation}
we identify
\begin{equation}
    Q_{\ell} \;=\;\left.\frac{1}{\ell!}\,\frac{{\rm d}^\ell}{{\rm d}z^\ell}\,\log t(z)\right|_{z=0},
    \qquad \ell=0,1,2\dots,
\end{equation}
so that knowledge of the \(R\)\nobreakdash-matrix determines \(t(z)\) and hence all conserved charges \(Q_\ell\).  

\subsection{Algebraic Bethe ansatz}
Rather than constructing each conserved charge individually, one can diagonalize the transfer matrix:
\begin{equation}
t(z)\ket{\Lambda}\;=\; \Lambda(z)\ket{\Lambda},
\end{equation}
where \(\ket{\Lambda}\) is an eigenvector of \(t(z)\) and \(\Lambda(z)\) its eigenvalue.  The eigenvalues of all charges 
\(Q_{n+1}\) then follow from the expansion
\begin{equation}
    \log\Lambda(z)\;=\;\sum_{n=0}^\infty \Lambda^{(Q_{n+1})}\,z^n
    \quad\Rightarrow\quad
    \Lambda^{(Q_{n+1})}
    = 
    \left.\frac{1}{n!}\,\frac{{\rm d}^n}{{\rm d}z^n}\log\Lambda(z)\right|_{z=0}.
\end{equation}
This diagonalization procedure, known as the {\it algebraic Bethe ansatz} (ABA), is especially powerful: it circumvents the 
exponential growth of the Hamiltonian matrix in site number by exploiting the integrable structure encoded in \(t(z)\).  

We decompose the monodromy matrix \(\mathcal{T}_{\alpha}(z)\) as
\begin{equation}
    \mathcal{T}_{\alpha}(z)
    = \mathcal{L}_{\alpha,N}(z)\,\mathcal{L}_{\alpha,N-1}(z)\,\cdots\,\mathcal{L}_{\alpha,1}(z)
    = \begin{pmatrix}
        \mathcal{A}(z) & \mathcal{B}(z) \\
        \mathcal{C}(z) & \mathcal{D}(z)
    \end{pmatrix},
\end{equation}
where \(\mathcal{A},\mathcal{B},\mathcal{C},\mathcal{D}\) act on the physical Hilbert space \(\mathcal{H}^{\otimes N}\) (of dimension \(2^N\)).  The transfer matrix is then
\begin{equation}
    t(z) \;=\; \operatorname{tr}_{\alpha}\mathcal{T}_{\alpha}(z)
    \;=\;
    \mathcal{A}(z) + \mathcal{D}(z).
\end{equation}
Consider the ferromagnetic pseudo-vacuum
\begin{equation}
\ket{0}=\bigotimes_{j=1}^N \begin{pmatrix}1 \\ 0\end{pmatrix}_{\!j},
\end{equation}
which satisfies
\begin{equation}
    \mathcal{C}(z)\ket{0} = 0,\quad
    \mathcal{A}(z)\ket{0}= a(z)\ket{0},\quad
    \mathcal{D}(z)\ket{0}= d(z)\ket{0}.
\end{equation}
Only the operator \(\mathcal{B}(z)\) creates excitations on \(\ket{0}\).  We define the {\it Bethe states}
\begin{equation}
    \ket{\{z_i\}}
    = \mathcal{B}(z_1)\,\mathcal{B}(z_2)\,\cdots\,\mathcal{B}(z_m)\ket{0}.
\end{equation}
For generic parameters \(\{z_i\}\), these are off-shell states; imposing the Bethe equations on \(\{z_i\}\) ensures that 
\(\ket{\{z_i\}}\) becomes an eigenvector of \(t(z)\).  

We start from
\begin{equation}
    t(z)\ket{\Lambda(z_1,\dots,z_m)}
    = \bigl[\mathcal{A}(z) + \mathcal{D}(z)\bigr]\,
      \mathcal{B}(z_1)\,\mathcal{B}(z_2)\,\cdots\,\mathcal{B}(z_m)\ket{0}.
\end{equation}
Using the RTT commutation relations, one moves \(\mathcal{A}(z)\) and \(\mathcal{D}(z)\) past the \(\mathcal{B}(z_i)\) operators 
and evaluates their action on \(\ket{0}\).  One then obtains
\begin{align}
t(z)\mathcal{B}(z_1)\cdots\mathcal{B}(z_m)\ket{0} 
&= \Lambda\bigl(z;\{z_i\}\bigr)\,\mathcal{B}(z_1)\cdots\mathcal{B}(z_m)\ket{0}\\
&\hspace{-70pt}+\sum_{i=1}^m 
\Bigl[\mathcal{M}_i^{(A)}\bigl(z;\{z_j\}\bigr)
+\mathcal{M}_i^{(D)}\bigl(z;\{z_j\}\bigr)\Bigr]\,
\mathcal{B}(z_1)\cdots\widehat{\mathcal{B}(z_i)}\cdots\mathcal{B}(z_m)\ket{0},
\nonumber
\end{align}
where \(\widehat{\mathcal{B}(z_i)}\) indicates omission of the \(i\)-th \(\mathcal{B}\) operator.  The functions \(\mathcal{M}_i^{(A,D)}\) arise from commutators with \(\mathcal{A}\) and \(\mathcal{D}\).  

Requiring that all unwanted terms vanish determines the {\it Bethe roots} \(\{z_i\}\) through
\begin{equation}
\mathcal{M}_i^{(A)}\bigl(z;\{z_j\}\bigr) + \mathcal{M}_i^{(D)}\bigl(z;\{z_j\}\bigr)
= 0
\quad\text{for}\;\;i=1,\dots,m.
\end{equation}
When these {\it Bethe equations} are satisfied, 
the second line disappears and \(\ket{\{z_i\}}\) becomes an eigenstate of the 
transfer matrix with eigenvalue \(\Lambda(z;\{z_i\})\).  

\section{Richardson--Gaudin models and Bethe equations}
\label{Ch3}
\subsection{The Richardson model}
We consider a system of fermions interacting via a reduced BCS (rBCS) Hamiltonian of the form:
\begin{equation}
H_{\rm rBCS} 
=\sum_{j,\sigma}\epsilon_j c_{j\sigma}^\dagger c_{j\sigma}-g\sum_{i,j} c_{i+}^\dagger c_{i-}^\dagger c_{j-} c_{j+},
\end{equation}
where \(\epsilon_j\) denotes the single particle energy level \(j\) ($j=1,\ldots,L$), \(g > 0\) is the coupling constant, 
and \(c_{j\sigma}^\dagger\), \(c_{j\sigma}\) are fermionic creation and annihilation operators with 
\(\sigma=+,-\) labeling time-reversed states.

When \(g=0\), electrons occupy single particle levels independently. A Cooper pair consisting of two electrons in time-reversed 
states \(\ket{\pm}\) contributes an energy \(2\epsilon_j\). 
For a general system with a given set \(\mathcal{S}\) of $N$ electrons, 
one can decompose the Hilbert space into two subspaces according to the underlying pairing structure. 
Specifically, let us denote (i) a paired subset \({\cal U}\subset{\cal S}\), with \(2M\) 
electrons forming \(M\) Cooper pairs; (ii) a blocked subset \({\cal B}={\cal S}\setminus{\cal U}\), consisting of 
\(K\!=\!N-2M\) unpaired electrons.
This decomposition of the Hilbert space allows us to restrict the dynamics to the paired sector \(\mathcal{U}\), where the 
reduced BCS Hamiltonian acts nontrivially. The corresponding state vector can be factorized as
\begin{equation}
\ket{M, K} = \prod_{i\in{\cal B}} c_{i\sigma}^\dagger\ket{\Psi_M}_{\cal U},
\end{equation}
where \(\ket{\Psi_M}_{\cal U}\) describes the wavefunction of the \(M\) Cooper pairs, and 
the product over \(\mathcal{B}\) creates the blocked single particle excitations. Since the singly occupied states do not 
participate in pairing, their total energy is simply additive:
\begin{equation}
\mathbb{E}_{\cal B}=\sum_{i\in{\cal B}}\epsilon_i=\sum_{i=1}^{K}\epsilon_i\equiv\mathbb{E}(K).
\end{equation}
By focusing on the subspace associated with \(\mathcal{U}\), one can solve the eigenvalue problem for the paired sector 
independently, using the exact Bethe ansatz methods described in the following subsections.

\paragraph{Hard--core boson representation.}
We define pair creation and annihilation operators as
\begin{equation}
b_j^\dagger = c_{j+}^\dagger c_{j-}^\dagger, \qquad b_j = c_{j-} c_{j+},
\end{equation}
which obey the following commutation relations:
\begin{equation}
b_j^{\dagger 2} = 0, \qquad [b_i, b_j^\dagger] = \delta_{ij}(1 - 2b_j^\dagger b_j), \qquad [b_j^\dagger b_j, b_k^\dagger] = 
\delta_{jk} b_j^\dagger
\end{equation}
(see appendix \ref{AppB1}).
These reflect the fermionic nature of the underlying electrons and define the so-called \emph{hard-core boson} algebra.

In this representation, the Hamiltonian acting on the paired subspace \({\cal U}\) becomes:
\begin{equation}\label{HU}
H_{\cal U}=\sum_{j\in{\cal U}}2\epsilon_j b_j^\dagger b_j-g\sum_{i,j\in{\cal U}} b_i^\dagger b_j.
\end{equation}
The factor of 2 in the Hamiltonian appears because each Cooper pair is made of two electrons. When there is no interaction 
(\(g=0\)), both electrons occupy the same energy level \(\epsilon_j\). Each one contributes \(\epsilon_j\) to the total energy, 
so the pair together contributes \(2\epsilon_j\). For this reason, we treat the energy level of a Cooper pair as \(2\epsilon_j\).

Indeed, let us start from the single particle energy term for level \(j\):
\begin{equation}
\sum_{\sigma} \epsilon_j c_{j\sigma}^\dagger c_{j\sigma} 
=\epsilon_j \left( c_{j+}^\dagger c_{j+} + c_{j-}^\dagger c_{j-} \right).
\end{equation}
We now employ the definition of the pair creation operator \( b_j^\dagger = c_{j+}^\dagger c_{j-}^\dagger \), and define the pair 
number operator as \(b_j^\dagger b_j\). 
In the subspace where level \(j\) is either empty or doubly occupied by a Cooper pair, we have:\footnote{ 
If level \(j\) is doubly occupied, then \(c_{j+}^\dagger c_{j+}=1\) and \( c_{j-}^\dagger c_{j-}=1\), so
\(\epsilon_j (1 + 1) = 2\epsilon_j\). 
In this case, \( b_j^\dagger b_j = 1 \), and thus the right-hand side is \( 2\epsilon_j b_j^\dagger b_j \).
If level \(j\) is empty, all number operators vanish, and so does \( b_j^\dagger b_j \), making both sides zero.
Therefore, the equality
\[
\epsilon_j\left(c_{j+}^\dagger c_{j+} + c_{j-}^\dagger c_{j-}\right) = 2\epsilon_j b_j^\dagger b_j
\]
holds when restricted to the subspace where single occupancy does not occur.}
\begin{equation}
c_{j+}^\dagger c_{j+} + c_{j-}^\dagger c_{j-} = 2 b_j^\dagger b_j,
\end{equation}
and thus
\begin{equation}
\sum_{\sigma} \epsilon_j c_{j\sigma}^\dagger c_{j\sigma} = 2\epsilon_j b_j^\dagger b_j.
\end{equation}
This step holds under the assumption that the Hilbert space is restricted to states where each energy level is either unoccupied 
or occupied by a full Cooper pair. That is, singly occupied states are excluded. This is the standard assumption in the reduced 
BCS model, which focuses solely on the dynamics of paired electrons. Outside of this subspace, the relation does not hold.

\paragraph{Bethe ansatz and Richardson equations.}
We seek eigenstates of the form:
\begin{equation}
\ket{\Psi_M}_{\cal U}=\prod_{\nu=1}^M B_\nu^\dagger\ket{0} 
\qquad \text{with} \qquad B_\nu^\dagger=\sum_{j\in{\cal U}} 
\frac{b_j^\dagger}{2\epsilon_j - E_\nu},
\end{equation}
where \(E_\nu\) is a spectral parameter associated with the \(\nu\)-th Cooper pair. The total energy of the system is then
\begin{equation}
\mathbb{E}(M)=\sum_{\nu=1}^{M}E_\nu.
\end{equation}
Introducing the collective operator
\begin{equation}
B_0^\dagger=\sum_{j\in{\cal U}} b_j^\dagger,
\end{equation}
we rewrite the Hamiltonian (\ref{HU}) as
\begin{equation}
H_{\cal U}=\sum_{j\in{\cal U}}2\epsilon_j b_j^\dagger b_j - g B_0^\dagger B_0.
\end{equation}
Applying \(H_{\cal U}\) to the ansatz state and simplifying using algebra detailed in appendix~\ref{AppB2}, we find:
\begin{align}\label{Prop1}
H_{\cal U}\ket{\Psi_M}_{\cal U} &=\mathbb{E}(M)\ket{\Psi_M}_{\cal U} 
+\sum_{\nu = 1}^{M}\left\{1-g\sum_{j\in{\cal U}}\frac{1}{2\epsilon_j - E_\nu} 
+ 2g\sum_{\mu\ne\nu}\frac{1}{E_\mu - E_\nu}\right\} \nonumber \\
&\qquad\times B_0^\dagger\prod_{\eta\ne\nu}B_\eta^\dagger\ket{0}.
\end{align}
For \(\ket{\Psi_M}_{\cal U}\) to be a true eigenstate, the second term must vanish. 
This yields a set of coupled nonlinear equations, known as the \textit{Richardson equations}:
\begin{equation}\boxed{
\frac{1}{g} - \sum_{j=1}^{L}\frac{1}{2\epsilon_j - E_\nu}+\sum_{\mu=1(\ne\nu)}^{M} 
\frac{2}{E_\mu-E_\nu}=0,\;\;\;\;\forall\;\nu = 1,\ldots,M.}\label{REq}
\end{equation}
These equations determine the allowed values of \(E_\nu\), which in turn define the exact energy spectrum and eigenstates of the 
reduced BCS Hamiltonian.

\subsection{Gaudin spin models and the conserved charges of the Richardson model}
There exists a profound connection between the Richardson reduced BCS model 
and a class of quantum integrable spin systems ({\it spin chains}) 
known as \textit{Gaudin models}~\cite{Gaudin1976, Gaudin2014}. 
These models are constructed from a given Lie algebra---typically \(\mathfrak{sl}
(2,\mathbb{C})\)---but the formalism extends naturally to other semisimple Lie algebras. A key feature of Gaudin models is the 
existence of mutually commuting Hamiltonians derived from the algebraic structure, governing the integrable dynamics of the 
system.

Let \(\mathfrak{g}\) be a semisimple Lie algebra of dimension \(d\), and consider \(L\) distinct complex parameters 
\(z_i\in\mathbb{C}\). The Gaudin Hamiltonians are defined as
\begin{equation}\label{HGG}
H_{\mathfrak{g},i} = \sum_{j \neq i} \sum_{a=1}^{d} \frac{I_a^{(i)} I^{a(j)}}{z_i - z_j},
\end{equation}
where \(\{I_a\}\) is a basis of \(\mathfrak{g}\), and \(\{I^a\}\) is its dual with respect to an invariant scalar product 
(typically the Killing form). These Hamiltonians act on the tensor product space
\[
\mathcal{V}_{\boldsymbol{\lambda}}\equiv\mathcal{V}_{\lambda_1} \otimes \cdots \otimes \mathcal{V}_{\lambda_L},
\]
where each \(\mathcal{V}_{\lambda_i}\) is a finite-dimensional irreducible representation of \(\mathfrak{g}\), labeled by a 
dominant weight \(\lambda_i\). The operator \(I_a^{(i)}\) acts nontrivially on the \(i\)-th tensor component and trivially on all 
others.

A central problem in the theory of Gaudin models is the simultaneous diagonalization of the commuting Hamiltonians 
\(H_{\mathfrak{g},i}\), 
which can be achieved via the algebraic Bethe ansatz originally introduced by Gaudin.

\paragraph{The \(\boldsymbol{\mathfrak{su}(2)}\) Gaudin model and the Richardson equations.}
In the context of the Richardson model, the relevant Gaudin model is associated with the \(\mathfrak{su}(2)\) algebra. The 
Hamiltonians take the form
\begin{equation}\label{Gaudin}
H_{\mathfrak{su}(2),i} = \sum_{j \neq i} \frac{1}{\epsilon_i - \epsilon_j}
\left[ S_i^z S_j^z + \frac{1}{2}(S_i^+ S_j^- + S_i^- S_j^+) \right] 
\equiv\sum_{j \neq i} \frac{\mathbf{S}_i \cdot \mathbf{S}_j}{\epsilon_i - \epsilon_j},
\end{equation}
where the \(\mathfrak{su}(2)\) generators are realized via hard-core boson operators:
\[
S_j^+ = b_j^\dagger, \quad S_j^- = b_j, \quad S_j^z = \frac{1}{2} - \hat{N}_j,
\]
with \(\hat{N}_j = b_j^\dagger b_j\) as the pair number operator. These Hamiltonians can be diagonalized using the 
Richardson--Bethe ansatz. The corresponding energy eigenvalues are
\begin{equation}\label{EGaudin}
\mathbb{E}_{\mathfrak{su}(2)}(M) = \sum_{\nu=1}^{M} E_\nu,
\end{equation}
where the parameters \(E_\nu\) satisfy the Richardson equations (\ref{REq}) in the limit \(g\to\infty\).
\begin{figure}[h!]
\centering
\includegraphics[width=0.5\textwidth]{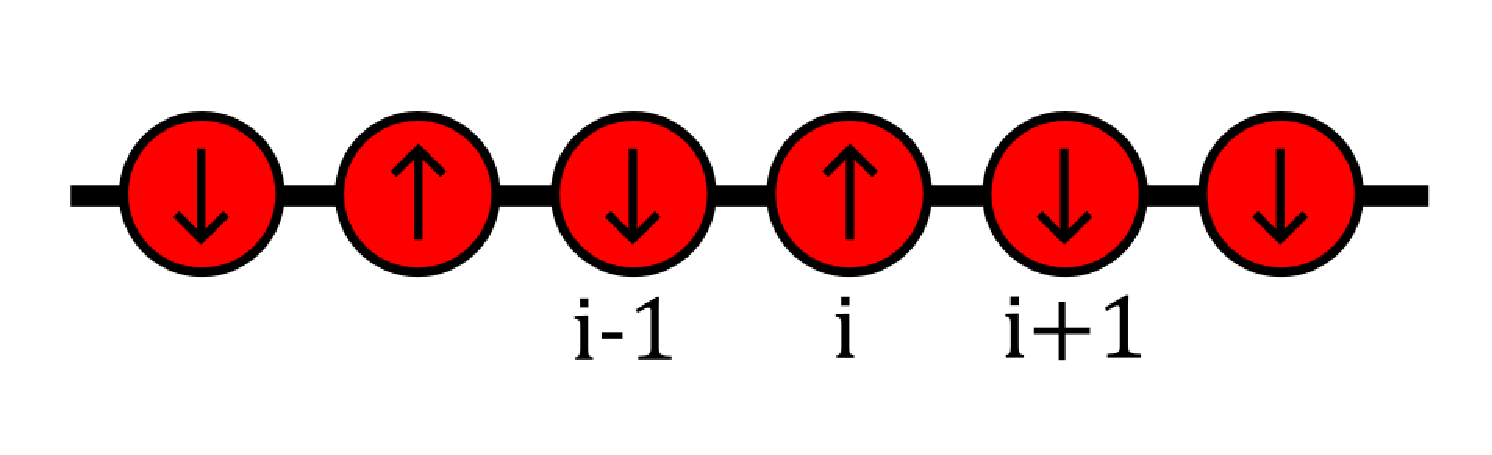}
\caption{
The \(\mathfrak{su}(2)\) Gaudin model describes a one-dimensional quantum spin system with long-range, position-dependent 
interactions. Each site \(i\) in the chain hosts a spin-\(\tfrac{1}{2}\) particle, represented by the generators 
\(\mathbf{S}_i=(S_i^+, S_i^-, S_i^z)\) of the Lie algebra \(\mathfrak{su}(2)\). The interactions between spins are governed by a 
set of mutually commuting Hamiltonians (\ref{Gaudin}), 
in which the coupling strengths depend on the parameters \(\epsilon_i\), which can be 
interpreted as position-like inhomogeneity parameters assigned to each site. These parameters introduce non-uniform interaction 
strengths of the form \(1/(\epsilon_i - \epsilon_j)\), leading to an integrable deformation of the uniform spin chain.
In this framework, the Bethe roots \(E_\nu\) arise as solutions to the Richardson (Bethe ansatz) equations in the limit 
\(g\to\infty\) and carry a clear physical meaning: each \(E_\nu\) corresponds to a collective excitation of the spin system. 
The total energy of an eigenstate is obtained as the sum over all Bethe roots (\ref{EGaudin}) 
where \(M\) is the number of such collective excitations.}
\label{fig:spin_chain}
\end{figure}

\paragraph{Conserved quantities and the rBCS Hamiltonian.}
The rational Gaudin Hamiltonians \eqref{Gaudin} naturally appear in the definition of conserved quantities of the rBCS model, 
known as the Cambiaggio--Rivas--Saraceno (CRS) integrals of motion~\cite{Cambiaggio:1997vz}:
\begin{equation}\label{Ri}
R_i = -S_i^z - gH_{\mathfrak{su}(2),i}
= -S_i^z - g \sum_{j \neq i} \frac{\mathbf{S}_i \cdot \mathbf{S}_j}{\epsilon_i - \epsilon_j}, \qquad i = 1, \ldots, L.
\end{equation}
These operators commute with each other and with the rBCS Hamiltonian. 
This allows the Hamiltonian to be re-expressed as~\cite{Sierra:1999mp,Cambiaggio:1997vz}:
\begin{align}
H_{\rm rBCS} &= H_{\rm XY} + \sum_{j=1}^{L} \epsilon_j + g\Big(\frac{1}{2}L-M\Big), \label{HR} \\
H_{\rm XY} &= \sum_{j=1}^{L} 2\epsilon_j R_j + g \Big(\sum_{j=1}^{L}R_j\Big)^2-\frac{3}{4}gL.
\end{align}
Hence, an alternative strategy to find the spectrum of \(H_{\rm rBCS}\) is to solve the eigenproblems for the quantum 
integrals of motion \(R_i\).

\paragraph{Connection to the central spin model.}
Each CRS operator \(R_i\) can be interpreted as a Hamiltonian of the \textit{central spin model} (or \textit{Gaudin 
magnet}), where a central spin \(\mathbf{S}_0\) interacts with a bath of \(L\) surrounding spins \(\mathbf{S}_j\):
\begin{equation}\label{Hcs}
H_{\rm cs} \equiv BR_{i=0} = -B S_0^z + \sum_{j=1}^{L} A_j \mathbf{S}_0 \cdot \mathbf{S}_j.
\end{equation}
The magnetic field \(B\) and inhomogeneous couplings \(A_j\) are parametrized to match the Richardson model:
\[
B = -\frac{2}{g}, \qquad A_j = \frac{2}{\epsilon_0 - \epsilon_j}.
\]
When \(\epsilon_0 = 0\), the parameters \(\epsilon_j\) coincide with the energy levels of the rBCS model.

\paragraph{Eigenvalues of the conserved charges and the Yang--Yang function.}
A closed form expression for the eigenvalues \(\lambda_i\) of the conserved charges 
\(R_i\) was obtained by 
Sierra~\cite{Sierra:1999mp} using techniques from two-dimensional conformal field theory (see subsection~\ref{Ad2dCFT}):
\begin{equation}\label{Li}
\lambda_i = \frac{g}{2} \left. \frac{\partial \mathcal{W}^{\rm R}_{\rm crit}({\bf z}, {\bf E}^{\rm c})}{\partial z_i} 
\right|_{z_i = 2\epsilon_i}
= -\frac{1}{2} + g\left( \sum_{\nu=1}^{M} \frac{1}{2\epsilon_i - E_\nu^{\rm c}} 
- \frac{1}{4} \sum_{j \neq i}^{L} \frac{1}{\epsilon_i-\epsilon_j}\right).
\end{equation}
Here, \({\bf E}^{\rm c} = \{E_1^{\rm c}, \ldots, E_M^{\rm c}\}\) is a solution of the Richardson equations, and 
\(\mathcal{W}^{\rm R}_{\rm crit}\) is the critical value of the YY function:
\begin{align}\label{U}
\mathcal{W}^{\rm R}({\bf z}, {\bf E}) &= -\sum_{i<j}^{L} \log(z_i - z_j) - 4\sum_{\nu < \mu}^{M} \log(E_\nu-E_\mu)\nonumber\\
&\quad + 2\sum_{i=1}^{L} \sum_{\nu=1}^{M} \log(z_i - E_\nu) + \frac{1}{g}\left(-\sum_{i=1}^{L}z_i 
+2 \sum_{\nu=1}^{M}E_\nu\right).
\end{align}
The conditions \(\partial\mathcal{W}^{\rm R}/\partial E_\nu = 0\), $\nu=1,\ldots,M$ 
recover the Richardson equations, confirming that \(\mathcal{W}^{\rm R}\) acts as a potential function for the Bethe ansatz 
equations.

Thus, an alternative formulation of the Richardson--Gaudin spectral problem involves the YY function (\ref{U})
(or functional, if the $E_\nu$ are interpreted as functions of the coupling 
$g$), whose stationary conditions reproduce the Richardson equations. This variational perspective 
not only provides a powerful numerical target---minimizing \(\mathcal{W}^{\rm R}\) 
via gradient-based methods\footnote{See the conclusions for further details.}---but also underlies an elegant 
electrostatic analogy: the rapidities \(E_\nu\) behave like unit ``charges'' on the complex plane, repelling one another and 
attracted to fixed ``background'' charges at the single particle energies \(\epsilon_j\).  

\subsection{Realization via 2D conformal field theory}
\label{Ad2dCFT}
The exact Richardson solution admits an interpretation in terms of conformal blocks of a two-dimensional conformal field 
theory.  In particular, to solve the eigenproblem for the Richardson conserved charges, Sierra \cite{Sierra:1999mp} showed 
that the Knizhnik--Zamolodchikov (KZ) equation for the $\widehat{\mathfrak{su}(2)}_{k}$ WZW block,
\begin{equation}
\biggl(\kappa\partial_{z_i}-\sum_{\substack{j=1 \\ j\neq i}}^{L+1}
\frac{\mathbf{S}_i\!\cdot\!\mathbf{S}_j}{z_i - z_j}\biggr)\psi^{\rm WZW}(z_1,\ldots,z_{L+1}) \;=\; 0,
\qquad
\kappa = \frac{k+2}{2},
\end{equation}
is entirely equivalent to the eigenvalue equation:
\begin{equation}\label{EigenR}
\frac{1}{2g}\,R_i\,\psi \;=\; -\,\kappa\,\partial_{z_i}\psi,
\qquad
\psi^{\rm WZW} \;=\;\exp\bigl[\,(2g\kappa)^{-1}\,H_{\rm XY}\bigr]\,\psi.
\end{equation}
Here, $\psi\equiv\psi^{\rm CG}_{\mathbf m}(\mathbf z)$ denotes a perturbed WZW conformal block in the free field (Coulomb gas) 
representation. More precisely, $\psi^{\rm CG}_{\bf m}({\bf z})$ consists of 
\begin{itemize}
\item[i.] the $\widehat{\mathfrak{su}(2)}_{k}$ WZW chiral primary fields
$\Phi_{m}^{j}(z)=\left(\gamma(z)\right)^{j-m}V_{\alpha}(z)$ 
built out of the $\gamma$-field of the $\beta\gamma$-system and Virasoro chiral vertex operators 
$V_{\Delta_{\alpha}}(z)$, $\Delta_{\alpha}=\alpha(\alpha-2\alpha_0)=j(j+1)/(k+2)$
represented as normal ordered exponentials of the free field $\phi(z)$.\footnote{Here, 
$\alpha=(k+2)^{-\frac{1}{2}}j=-2\alpha_0 j$.} 
\item[ii.] WZW screening charges:
$$
{\sf Q}=\oint\limits_{C}\frac{{\rm d}E}{2\pi i}\,S(E)\,,
\quad\quad
S(E)=\beta(E)V_{2\alpha_0}(E)\,;
$$
\item[iii.] the operator
$$
V_{g}=\exp\!\left[-\frac{i\alpha_0}{g}\oint\limits_{C_g}{\rm d}z\,z\partial_z\phi(z)\right],
$$
which breaks conformal invariance. 
\end{itemize}

Within this realization to every energy level 
$z_i=2\epsilon_i$ corresponds the field $\Phi_{m_i}^{j}(z_i)$
with the spin $j=\frac{1}{2}$ and the ``third component'' $m_i=\frac{1}{2}$
(or $m_i=-\frac{1}{2}$) if the corresponding energy level is empty (or occupied) by a pair of fermions.
Integration variables $E_{\nu}$ in screening operators are the Richardson parameters.
The operator $V_{g}$ implements the coupling $g$ and is a source of the term $\frac{1}{g}$ in 
the Richardson equations (\ref{REq}). 

After ordering, $\psi^{\rm CG}_{\bf m}({\bf z})$ has a structure of a multidimensional contour integral,
\begin{equation}\label{RCG}
\psi^{\rm CG}_{\bf m}({\bf z})=\oint\limits_{C_1}{\rm d}E_{1}\ldots
\oint\limits_{C_M}{\rm d}E_{M}\,
\psi^{\beta\gamma}_{\bf m}({\bf z},{\bf E})\,{\rm e}^{-\alpha_{0}^{2}{\cal W}^{\rm R}({\bf z},{\bf E})},
\end{equation}
where ${\cal W}^{\rm R}({\bf z},{\bf E})$ is given by (\ref{U}).
In the limit $\alpha_{0}\to\infty$ $\Leftrightarrow$ $k\to -2$ $\Leftrightarrow$ $\kappa\to 0$, this integral
can be calculated using the saddle point method. 
The stationary solutions are then given by the solutions of the Richardson equations.
After all one gets 
\begin{equation}\label{AsympR}
\psi^{\rm CG}_{\bf m}({\bf z})\;\sim\;\psi^{\rm R}\;
{\rm e}^{-\alpha_{0}^{2}{\cal W}^{\rm R}_{\rm crit}({\bf z},{\bf E}^{\rm c})}
\quad{\rm for}\quad \alpha_{0}\to\infty, 
\end{equation}
where $\psi^{\rm R}\equiv\psi^{\beta\gamma}_{\bf m}({\bf z},{\bf E}^{\rm c})$ is the Richardson wave function.
Using this asymptotic limit to the equation (\ref{EigenR}) one obtains
$R_{i}\psi^{\rm R}=\lambda_{i}\psi^{\rm R}$ for $\kappa\to 0$, where $\lambda_{i}$ are given by (\ref{Li}).

\paragraph{Yang--Yang function via Gaiotto--Witten irregular conformal blocks.}
The subsequent points represent novel findings beyond those in \cite{Sierra:1999mp}.
The saddle-point asymptotic in \eqref{AsympR} is exactly the large-\(c\) limit of the conformal 
blocks.\footnote{Here, $c=3-12\alpha_{0}^{2}=3k/(k+2)$.} 
More generally, in this limit the saddle values of the Coulomb gas integrals---i.e., the critical values 
\(\mathcal{W}_{\rm crit}\) of the YY functions---coincide with the exponentiated classical conformal blocks, 
which in turn admit independent power series expansions.
Since finding \({\cal W}_{\rm crit}\) normally requires solving the Bethe (saddle-point) equations, one can instead 
approximate it by truncating the series for the classical block.
In subsection~\ref{EYY}, we present a concrete example: the Dotsenko--Fateev integral representation of the Virasoro four-point 
conformal block on the sphere. In the classical (saddle-point) limit, its saddle-point equations reduce to a Gaudin-type system.

The construction of the function ${\cal W}^{\rm R}_{\rm crit}$ in \cite{Sierra:1999mp}---which both generates the full spectrum 
of conserved charges in the Richardson model and solves the eigenvalue problem for the Gaudin magnet (central spin) 
Hamiltonian---reveals a deep link between quantum integrability, conformal symmetry, and the physics of Richardson--Gaudin 
models. However, because this construction explicitly breaks conformal invariance, one cannot directly invoke standard 2D CFT 
machinery (e.g., factorization over a complete basis of intermediate states) to compute conformal blocks or their power series 
coefficients---though a perturbed CFT approach might still apply. Remarkably, one can reformulate the YY function for 
RG models in a manner that preserves conformal symmetry and permits the use of all standard techniques of 
two-dimensional conformal field theory.

The YY function (\ref{U}) also emerges in a setting far removed from conventional quantum integrability. In their study of a 
four-dimensional gauge-theoretic approach to the Jones polynomial, Gaiotto and Witten \cite{GW} 
recognized that the relevant saddle-point equations coincide with the Bethe ansatz system (\ref{REq}). Moreover, Gaiotto 
and Witten exhibited a free field realization of the associated YY function in terms of irregular, degenerate Virasoro conformal 
blocks. Their construction uses the usual degenerate vertex operators and screening charges together with a rank-one irregular 
insertion at infinity, which creates the corresponding irregular state from the vacuum.\footnote{See also recent 
analyses of these irregular conformal blocks in \cite{GH,HLR,GHLR}.}

Let us describe this construction in more detail. We denote by 
$V_\alpha(z)\!=:\!{\rm e}^{2\alpha\varphi(z)}\!:$
the Virasoro chiral vertex operator of ``momentum'' \(\alpha\in\mathbb{C}\), central charge
\[
c \;=\;1+6Q^2,\qquad Q=b+b^{-1},
\]
and conformal weight
\[
\Delta_\alpha \;=\;\alpha(Q-\alpha).
\]
The standard degenerate primaries occur at
\[
\alpha_{r,s}
\;=\;
\frac{1-r}{2}\,b \;+\;\frac{1-s}{2}\,b^{-1},
\qquad r,s\in\mathbb{Z}_{>0},
\]
with
\[
V_{r,s}(z)\;=\;:{\rm e}^{2\alpha_{r,s}\varphi(z)}:\,, 
\qquad
\Delta_{r,s}
\;=\;
\alpha_{r,s}(Q-\alpha_{r,s})
\;=\;
\Delta_0 \;-\;\frac{(r\,b + s\,b^{-1})^2}{4},
\]
where
\[
\Delta_0 \;=\;\frac{Q^2}{4}
\;=\;\frac{c-1}{24}.
\]
The Gaiotto--Witten conformal block is defined as the chiral correlator
\[\int\limits_{\Gamma}
\Big\langle I_{\alpha,\gamma}(\infty)\;\prod_{i=1}^L V_{-\frac{k_i}{2b}}(z_i)\;\prod_{\nu=1}^M V_{\frac{1}{b}}(E_\nu)
\Big\rangle\,{\rm d}E_1\ldots {\rm d}E_M,
\]
where the degenerate primaries 
\[
V_{-\tfrac{k_i}{2b}}(z_i)\equiv V_{1,k_i+1}(z_i),\quad k_i\in\mathbb{N}_{>0},
\]
are inserted at positions \(z_i\), and the operators
$V_{\tfrac{1}{b}}(E_\nu)$ are integrated over appropriate contours.
Crucially, the chiral correlator also includes a single rank-one irregular vertex operator
$$
I_{\alpha,\gamma}(z)\;\equiv\;
:{\rm e}^{2\alpha\varphi(z)}{\rm e}^{\gamma\partial\varphi(z)}:
$$
inserted at $z=\infty$.\footnote{Cf.~appendix \ref{Irr1}.}
Taken together, these ingredients define the irregular, degenerate conformal block, which can be represented as follows 
\cite{HLR}:
\begin{eqnarray}\label{FGW}
{\cal F}_{\Gamma}({\bf z},{\bf k})
=\int\limits_{\Gamma}\exp\!\left[-\frac{1}{b^2}{\cal W}^{\rm GW}
\!({\bf z},{\bf E},{\bf k})\right]{\rm d}E_1\ldots {\rm d}E_M,
\end{eqnarray}
where 
\begin{eqnarray}\label{W}
{\cal W}^{\rm GW}\!({\bf z},{\bf E},{\bf k}) 
&=&
\frac{1}{2}\sum\limits_{i<j}k_ik_j\log(z_i-z_j)
+2\sum\limits_{\nu<\mu}\log(E_\nu-E_\mu)\nonumber
\\
&-&	
\sum\limits_{i}\sum\limits_{\nu}k_i\log(E_\nu-z_i)
-\Lambda\left(-\frac{1}{2}\sum\limits_{i}k_i z_i+\sum\limits_{\nu}E_{\nu}\right).
\end{eqnarray}
In \eqref{FGW}, $\Gamma$ may be chosen to be any contour obtained by deforming the integration paths of the screening charges so 
as to ensure convergence of the integral.
In \eqref{W}, $\Lambda$ represents a constant that is directly proportional to the eigenvalue of the rank-one irregular state 
when acted upon by the Virasoro generator $L_{1}$.

Let us note that when we specialize to $\mathbf{k}=\{k_i\!=\!1\}_{i=1}^{L}$
the Gaiotto--Witten YY function
$\mathbb{W}(\mathbf{z},\mathbf{E})\!\equiv\!\mathcal{W}^{\rm GW}(\{z_i\},\{E_\nu\},\{1\})$
coincides with the Richardson YY function $\mathcal{W}^{\rm R}(\mathbf{z},\mathbf{E})$ 
defined in \eqref{U}. More precisely, one finds
$\mathcal{W}^{\rm R}(\mathbf{z},\mathbf{E})=
-2\mathbb{W}(\mathbf{z},\mathbf{E})
+2\pi iLM$,
provided that the identification
$\frac{1}{g}=\Lambda$
is made between the Richardson coupling $g$ and the ``irregularity'' parameter $\Lambda$.
Therefore, the extremal value $\mathbb{W}_{\rm crit}(\mathbf{z},\mathbf{E}^{\rm c})$ encodes 
the conserved charges of the reduced BCS model.  Indeed, since
$\partial_{E_\nu}\mathcal{W}^{\rm R}=-2\partial_{E_\nu}\mathbb{W}$
and
$\partial_{z_i}\mathcal{W}^{\rm R}=-2\partial_{z_i}\mathbb{W}$,
it follows from \eqref{Li} that
$$
\lambda_i
\;=\;
-\frac{1}{\Lambda}\,
\frac{\partial\,\mathbb{W}_{\rm crit}(\mathbf{z},\mathbf{E}^{\rm c})}
{\partial z_i}
\Bigg|_{\,z_i=2\epsilon_i}.
$$

The key question is whether this correspondence offers genuinely new insights.  As noted above, expressing the critical 
Yang--Yang function ${\cal W}^{\rm R}_{\rm crit}$ in terms of irregular Gaiotto--Witten blocks 
immediately grants access to the full toolkit of 2D CFT. 
In particular, these blocks are degenerate---hence they satisfy BPZ \cite{BPZ} differential equations arising from the decoupling 
of null states. By taking the classical (large-$c$, $b\to 0$) limit of those BPZ equations and performing a saddle-point 
analysis, one recovers precisely the equations governing ${\cal W}^{\rm R}_{\rm crit}$, 
in direct analogy to how BPZ equations control classical 
conformal blocks. This observation opens the door to analytic methods for studying 
${\cal W}^{\rm R}_{\rm crit}$.

\section{Eigenvalue--based reformulation}
\label{Ch4}
\subsection{Eigenvalue--based method}
This section begins by rewriting the Richardson equations~(\ref{REq}) using a new notation and a different indexing convention:
\begin{equation}\label{eq:4_ebm-rich_eqs}
\boxed{\frac{1}{g}-\sum_{\alpha=1}^{L}\frac{1}{\epsilon_\alpha-\lambda_j}
+\sum_{\substack{k = 1\\k\neq j}}^{M}\frac{2}{\lambda_k - \lambda_j}=0,
\qquad\forall\;j = 1,\ldots,M.}
\end{equation}
This reformulation aligns with the conventions widely adopted in the literature, 
particularly in the work of Faribault \textit{et al.}~\cite{Faribault2011}, 
which serves as a primary reference throughout this section.
The set of Bethe roots (\textit{rapidities}), denoted in the previous section as \(\{E_\nu\}\), 
is here denoted by \(\{\lambda_j\} \). In addition, in equation (\ref{eq:4_ebm-rich_eqs}) we performed a change of variable 
$2\epsilon\mapsto\epsilon$: originally $2\epsilon$ represented the pair energy (with $\epsilon$ the single electron energy in the 
non-interacting limit), whereas now $\epsilon$ denotes the total energy of the Cooper pair.

Directly solving equations (\ref{eq:4_ebm-rich_eqs}) with standard numerical algorithms for nonlinear systems is highly 
challenging due to the presence of singularities that occur when the rapidities $\lambda_j$ approach $\epsilon_\alpha$.
To overcome this difficulty, we employ an alternative approach in which such 
singularities are avoided. As already mentioned, this section is based on the work~\cite{Faribault2011}, 
and it outlines techniques for solving the Richardson equations~(\ref{eq:4_ebm-rich_eqs}) using the so-called 
\textit{eigenvalue-based method}.\footnote{See also \cite{Faribault2012}.}

First, let us consider a polynomial of degree \(M\) in the form
\begin{equation}\label{Bax}
P(z) = \prod_{j=1}^{M}\left(z-\lambda_j\right),
\end{equation}
where \(z\) is a complex variable. The set \(\{\lambda_j\}\) of roots of this polynomial corresponds to the set of rapidities, 
which are the solutions of the Richardson equations.

The derivative of \(P(z)\) with respect to \(z\) is given by
\begin{equation}
P'(z) = \sum_{j=1}^{M} \left[ \prod_{\substack{k=1\\k\neq j}}^{M} (z - \lambda_k) \right].
\end{equation}

We now define a function \(\Lambda(z)\) as
\begin{equation}
\Lambda(z) = \frac{P'(z)}{P(z)} = \sum_{j=1}^{M} \frac{1}{z-\lambda_j}.
\end{equation}
Its derivative is then given by
\begin{equation}
\Lambda'(z) = -\sum_{j=1}^{M} \frac{1}{(z - \lambda_j)^2}.
\end{equation}
Clearly, \(\Lambda(z)\) satisfies a Riccati-type differential equation~\cite{Faribault2011}:
\begin{equation}
\frac{{\rm d}\Lambda(z)}{{\rm d}z} + \Lambda^2(z) = 
-\sum_{j=1}^{M} \frac{1}{(z - \lambda_j)^2} + \sum_{j,k=1}^{M} \frac{1}{(z - \lambda_j)(z - \lambda_k)}.
\end{equation}
It can also be shown that if the set \(\{\lambda_j\}\) satisfies the Bethe equations, then the following identity 
holds~\cite{Faribault2011}:
\begin{equation}\label{eq:4_ebm-F_intro}
\Lambda'(z) + \Lambda^2(z) - \sum_{\alpha=1}^{M}\frac{2F(\lambda_{\alpha})}{\left(z-\lambda_{\alpha}\right)} = 0.
\end{equation}
In our case, the function $F(\lambda_{\alpha})$ will be given by
\begin{equation}
F(\lambda_{\alpha}) = -\sum_{i=1}^{L}\frac{A_i}{\left(\epsilon_i-\lambda_{\alpha}\right)}+\frac{B}{2g}\lambda_{\alpha}+ 
\frac{C}{2g}.
\end{equation}
The physical interpretation of the parameters \(\epsilon_j\) and \(g\) is model-dependent. In the case of the Richardson 
equations, the set \(\{\epsilon_j\}\) corresponds to the single particle energy levels, while \(g\) represents the 
coupling constant.

Substituting this form of \(F(\lambda_{\alpha})\) into equation~(\ref{eq:4_ebm-F_intro}) yields:
\begin{equation}
\Lambda'(z) + \Lambda^2(z) - \sum_{\alpha=1}^{M}\frac{2}{\left(z-\lambda_\alpha\right)}\left[-\sum_{i=1}^{L}\frac{A_i}
{\left(\epsilon_i-\lambda_{\alpha}\right)}+\frac{B}{2g}\lambda_{\alpha} + \frac{C}{2g}\right] = 0.
\end{equation}
In the limit $z\to\epsilon_j$ we thus get
\begin{align}
0 &= \Lambda'(\epsilon_j) + \Lambda^2(\epsilon_j) + \sum_{\alpha=1}^{M}\sum_{i=1}^{L}\frac{2A_i}{\left(\epsilon_j-
\lambda_{\alpha}\right)\left(\epsilon_i-\lambda_{\alpha}\right)} -\frac{B}{g}\sum_{\alpha=1}^{M}\frac{\lambda_{\alpha}}
{\left(\epsilon_j-\lambda_{\alpha}\right)}
\nonumber\\
&\hspace{90pt}-\frac{C}{g}\sum_{\alpha=1}^{M}\frac{1}{\left(\epsilon_j-\lambda_{\alpha}\right)}
\nonumber\\
&= \Lambda'(\epsilon_j) + \Lambda^2(\epsilon_j) + \sum_{\alpha=1}^{M}\left[\sum_{i\neq j}^{L}\frac{2A_i}{\left(\epsilon_i-
\lambda_{\alpha}\right)\left(\epsilon_j-\lambda_{\alpha}\right)}+\sum_{i=1}^{L}\frac{2A_i\delta_{ij}}{\left(\epsilon_i-
\lambda_{\alpha}\right)\left(\epsilon_j-\lambda_{\alpha}\right)}\right] 
\nonumber\\
&\hspace{90pt}-\sum_{\alpha=1}^{M}\frac{B\lambda_{\alpha}+C}{g\left(\epsilon_j-\lambda_{\alpha}\right)}
\nonumber\\
&=\Lambda'(\epsilon_j) + \Lambda^2(\epsilon_j) + \sum_{\alpha=1}^{M}\sum_{i\neq j}^{L}\frac{\epsilon_i-\epsilon_j}{\epsilon_i-
\epsilon_j}\frac{2A_i}{\left(\epsilon_i-\lambda_{\alpha}\right)\left(\epsilon_j-\lambda_{\alpha}\right)}  
\nonumber\\
&\hspace{90pt}+\sum_{\alpha=1}^{M}\frac{2A_j}{\left(\epsilon_j-\lambda_{\alpha}\right)\left(\epsilon_j-
\lambda_{\alpha}\right)} -\sum_{\alpha=1}^{M}\frac{B\lambda_{\alpha} + B\epsilon_j-B\epsilon_j+C}{g\left(\epsilon_j-
\lambda_{\alpha}\right)}
\nonumber\\
&= \Lambda'(\epsilon_j) + \Lambda^2(\epsilon_j) + \sum_{i\neq j}^{L}\frac{2A_i}{\epsilon_i-\epsilon_j}\sum_{\alpha=1}^{M}
\frac{\epsilon_i-\lambda_{\alpha} - \epsilon_j +\lambda_{\alpha}}{\left(\epsilon_i-\lambda_{\alpha}\right)\left(\epsilon_j-
\lambda_{\alpha}\right)} 
\nonumber\\
&\hspace{90pt}+2A_j\sum_{\alpha=1}^{M}\frac{1}{\left(\epsilon_j-\lambda_{\alpha}\right)^2}+ \sum_{\alpha=1}^{M}
\left[\frac{B\left(\epsilon_j-\lambda_{\alpha}\right)}{g\left(\epsilon_j-\lambda_{\alpha}\right)} - \frac{B\epsilon_j+C}
{g\left(\epsilon_j-\lambda_{\alpha}\right)}\right]  
\nonumber\\
&= \Lambda'(\epsilon_j) + \Lambda^2(\epsilon_j) +\sum_{i\neq j}^{L}\frac{2A_i}{\epsilon_i-\epsilon_j}\sum_{\alpha=1}^{M}
\left[\frac{\epsilon_i-\lambda_{\alpha}}{\left(\epsilon_i-\lambda_{\alpha}\right)\left(\epsilon_j-\lambda_{\alpha}\right)} - 
\frac{\epsilon_j-\lambda_{\alpha}}{\left(\epsilon_i-\lambda_{\alpha}\right)\left(\epsilon_j-\lambda_{\alpha}\right)}\right] 
\nonumber\\
&\hspace{90pt}+2A_j\left[-\Lambda'(\epsilon_j)\right]+\sum_{\alpha=1}^{M}\frac{B}{g}-\frac{B\epsilon_j+C}{g}
\sum_{\alpha=1}^{M}\frac{1}{\epsilon_j-\lambda_{\alpha}} 
\nonumber\\
&= \Lambda'(\epsilon_j)+\Lambda^2(\epsilon_j)+\sum_{i \neq j}^{L}\frac{2A_i}{\epsilon_i - \epsilon_j}\sum_{\alpha=1}^{M}
\left[\frac{1}{\epsilon_j-\lambda_{\alpha}}-\frac{1}{\epsilon_{i}-\lambda_{\alpha}}\right]  
\nonumber\\
&\hspace{90pt}-2A_j\Lambda'(\epsilon_j)+\frac{B}{g}M-\frac{B\epsilon_j+C}{g}\Lambda(\epsilon_j)  
\nonumber\\
&= \left(1-2A_j\right)\Lambda'(\epsilon_j)+\Lambda^2(\epsilon_j) +\sum_{i\neq j}^{L}\frac{2A_i\left[\Lambda(\epsilon_j)-
\Lambda(\epsilon_i)\right]}{\epsilon_i-\epsilon_j} + \frac{B}{g}M - \frac{B\epsilon_j+C}{g}\Lambda(\epsilon_j). 
\label{eq: 4_ebm-gen_sol}
\end{align}
In the pseudospin models we assume that $A_i = |s_i|\Omega_i$, where $|s_i|$ is the norm of the local spin degree of freedom, and $\Omega_i$ is an integer related to the degeneracy.

\subsection{Application to Richardson's equations}
Let us now examine the form of \( F(\lambda_{\alpha})\) and the Richardson equations (\ref{eq:4_ebm-rich_eqs}) cast in the form
\begin{equation}\label{REF}
-\sum_{\substack{k = 1\\k\neq j}}^{M}\frac{1}{\lambda_{k}-\lambda_{j}} = -\sum_{\alpha=1}^{L}\frac{1}{2(\epsilon_{\alpha}-
\lambda_j)} + \frac{1}{2g}.
\end{equation}
We observe that the right-hand side of equation~(\ref{REF}) 
takes the same form as \( F(\lambda_j) \) 
if we choose \( A_i = \frac{1}{2} \), \( B = 0 \), and \( C = 1 \).
Substituting these values into the general 
solution~(\ref{eq: 4_ebm-gen_sol}) yields:
\begin{equation}\label{eq: 4_ebm-quad1}
\Lambda^2(\epsilon_j) + \sum_{i \neq j}^{L}\frac{\Lambda(\epsilon_j) - \Lambda(\epsilon_i)}{\epsilon_i-\epsilon_j} +\frac{1}{g}
\Lambda(\epsilon_j) = 0,\text{\quad} j=1,\ldots,M,
\end{equation}
which is a set of $M$ quadratic equations for $\Lambda(\epsilon_j)$. 
For simplicity let us multiply (\ref{eq: 4_ebm-quad1}) by $g^2$, and substitute $\Lambda_j = g\Lambda(\epsilon_j)$. 
Then, one gets
\begin{equation}\label{eq: 4_ebm-quad2}
\Lambda_j^2 + g\sum_{i\neq j}^{L}\frac{\Lambda_j - \Lambda_i}{\epsilon_i - \epsilon_j} + \Lambda_j = 0,\text{\qquad} j = 1,\ldots,M.
\end{equation}
This set of equations can be solved numerically using a variety of methods, one of which is the Newton--Raphson method.

For the boundary conditions at $g=0$ we take
\begin{equation}\label{eq: 4_ebm-lambda_init}
    \Lambda_j = \begin{cases}
        &1\text{\quad if $\epsilon_j$ is occupied,}\\
        &0\text{\quad if $\epsilon_j$ is not occupied.}
    \end{cases}
\end{equation}
Solving (\ref{eq: 4_ebm-quad2}) up to desired precision gives us a set of $M$ values of $\Lambda_j$. 
After dividing by $g$, we get desired values of $\Lambda(\epsilon_j)$. We can use these values to find the coefficients of 
polynomials $P(z)$. If we write $P(z)$ and $P'(z)$ in the form
\begin{align}
P(z) = \sum_{m=0}^{m}P_{M-m}\,z^{m},\text{\qquad} P'(z) = \sum_{m=0}^{m}mP_{M-m}\,z^{m-1},
\end{align}
we have
\begin{equation}
\Lambda(z) = \frac{\sum\limits_{m=0}^{M}mP_{M-m}\,z^{m-1}}{\sum\limits_{m=0}^{M}P_{M-m}\,z^{m}}
\;\;\;\Rightarrow\;\;\;
\Lambda(\epsilon_j) = \frac{\sum\limits_{m=0}^{M}mP_{M-m}\epsilon_{j}^{m-1}}{\sum\limits_{m=0}^{M}P_{M-m}\epsilon_{j}^{m}},
\end{equation}
with $P_0=1$. It is worth noting that the index $M - m$ does not refer to the power of $z$, but rather to the order of the 
product over the roots $\lambda_j$. Given $L$ known values of $\Lambda(\epsilon_j)$, the problem can be reduced to a linear one. 
We obtain:
\begin{eqnarray}
\sum_{m=0}^{M}P_{M-m}\epsilon_j^{m}\Lambda(\epsilon_j) &=& \sum_{m=0}^{M}mP_{M-m}\epsilon_j^{m-1}\;\;\;\Rightarrow
\nonumber\\
\sum_{m=0}^{M-1}P_{M-m}\epsilon_j^{m}\Lambda(\epsilon_j) + P_{M-M}\epsilon_j^{M}\Lambda(\epsilon_j) &=& \sum_{m=0}^{M-1}mP_{M-m}\epsilon_j^{m-1} + MP_{M-M}\epsilon_j^{M-1}\;\;\;\Rightarrow
\nonumber\\
\epsilon_j^{M}\Lambda(\epsilon_j) - M\epsilon_j^{M-1} &=& \sum_{m=0}^{M-1}\left[m\epsilon_j^{m-1}P_{M-m} - \epsilon_j^{m}\Lambda(\epsilon_j)P_{M-m}\right]\;\;\;\Rightarrow
\nonumber\\
\label{eq: 4_ebm-lin_P}
\sum_{m=0}^{m-1}\left[m\epsilon_j^{m-1}-\epsilon_j^{m}\Lambda(\epsilon_j)\right]P_{M-m} &=& \epsilon_j^{M}\Lambda(\epsilon_j) - M\epsilon_j^{M-1},
\end{eqnarray}
which is a set of $L$ linear equations for $M-1$ coefficients of $P(z)$.

In the case where $M < L$ and the resulting system of equations is directly solvable, one can construct the polynomial $P(z)$ and 
determine its roots using the Laguerre method with polynomial deflation. This approach is particularly effective because the 
Laguerre method is guaranteed to converge to at least one (possibly complex) root. Polynomial deflation subsequently allows us to 
reduce the degree of the polynomial by factoring out the discovered root, thereby simplifying the root-finding problem. Repeating 
this procedure enables the systematic determination of all roots of the polynomial.

To determine the dependence of the rapidities $\{\lambda_j\}$ on the coupling constant $g$, we begin with the initial values of $
\Lambda_j$ given by equation~(\ref{eq: 4_ebm-lambda_init}). The first set of rapidities is computed for a small, non-zero value 
of $g$. In subsequent iterations, $g$ is incrementally increased by a small factor, and the previously obtained values of $
\Lambda_j$ are used as initial conditions for the next step. This iterative procedure allows for a controlled and continuous 
tracking of the evolution of the rapidities as a function of the coupling strength.

In the case where $M > L$, the system of equations given in~(\ref{eq: 4_ebm-lin_P}) becomes undetermined and cannot be solved 
directly. This corresponds to a degenerate scenario. To address this issue, the energy levels are extended into the complex 
domain such that each Cooper pair occupying a degenerate energy level $\epsilon_j$ is assigned to a distinct complex energy level 
$\varepsilon_j$, defined by
\begin{equation}\label{deg}
\varepsilon_j = \epsilon_j + i k, \qquad k = -\frac{d_j-1}{2},\ -\frac{d_j-1}{2}+1,\ldots,\frac{d_j-1}{2}-1,\ \frac{d_j-1}{2},
\end{equation}
where $d_j$ denotes the degeneracy of the energy level $\epsilon_j$ in the $g = 0$ limit. For example, if the level $\epsilon_j$
is occupied by three Cooper pairs, then $d_j = 3$, and the corresponding complex levels are $\varepsilon_j = \epsilon_j - i,\
\epsilon_j$, and $\epsilon_j + i$. This extension into the complex plane guarantees that the number of distinct $\varepsilon_j$
matches the number of Cooper pairs, ensuring solvability of the system. Moreover, the symmetric distribution of the complex
shifts about the real axis preserves the reality of the total energy.

\section{Numerical implementation}
\label{Ch5}
In this section, we present the details of the numerical implementation used to solve the Richardson equations. 
The primary goal of the program is to efficiently compute the rapidities for varying values of the coupling constant $g$, 
including both non-degenerate and degenerate cases. The implementation combines symbolic preprocessing with robust numerical 
algorithms, such as Newton--Raphson iteration and the Laguerre method with polynomial deflation, to ensure stability and 
convergence. 

The structure of the code was designed to be modular and flexible, allowing for straightforward extension to related models. This 
section outlines the overall architecture of the program, the key numerical methods employed, and practical considerations such 
as initialization strategies, handling of complex roots, and performance optimization.

We begin by characterizing the numerical methods mentioned above.

\subsection{LU decomposition}
Let us consider a system of \(N\) linear equations in matrix form:  
\begin{equation}\label{eq:4_num-LU1}
    \mathbf{A}\,\mathbf{x} = \mathbf{b},
\end{equation}
where \(\mathbf{A}\) is an \(N\times N\) matrix, \(\mathbf{b}\) is a given \(N\)-element column vector, and \(\mathbf{x}\) is the unknown \(N\)-element column vector that satisfies this equation.  Suppose that \(\mathbf{A}\) admits an LU factorization,  
\begin{equation}
    \mathbf{A} = \mathbf{L}\,\mathbf{U},
\end{equation}
where \(\mathbf{L}\) is a lower triangular matrix with entries \(\alpha_{ij}\) (\(\alpha_{ij}=0\) for \(j>i\)), and \(\mathbf{U}\) is an upper triangular matrix with entries \(\beta_{ij}\) (\(\beta_{ij}=0\) for \(j<i\)).  Then \eqref{eq:4_num-LU1} becomes
\begin{equation}
    \mathbf{L}\,\mathbf{U}\,\mathbf{x} = \mathbf{b}.
\end{equation}

If we set \(\mathbf{U}\,\mathbf{x} = \mathbf{y}\), then
\begin{equation}\label{eq:4_num-LU_y}
    \mathbf{L}\,\mathbf{y} = \mathbf{b}.
\end{equation}
Since \(\mathbf{L}\) is lower triangular, we solve \eqref{eq:4_num-LU_y} by forward substitution.  The first component is
\begin{equation}
    y_0 = \frac{b_0}{\alpha_{00}},
\end{equation}
and the remaining components follow iteratively as
\begin{equation}
    y_i = \frac{1}{\alpha_{ii}}\Bigl[b_i - \sum_{j=0}^{i-1}\alpha_{ij}\,y_j\Bigr],
    \quad i = 1,2,\dots,N-1.
\end{equation}
Having obtained \(\mathbf{y}\), we solve \(\mathbf{U}\,\mathbf{x} = \mathbf{y}\) by back substitution:
\begin{align}
    x_{N-1} &= \frac{y_{N-1}}{\beta_{N-1,N-1}},\\
    x_i &= \frac{1}{\beta_{ii}}\Bigl[y_i - \sum_{j=i+1}^{N-1}\beta_{ij}\,x_j\Bigr],
    \quad i = N-2,N-3,\dots,0.
\end{align}

\subsection{Newton--Raphson method}
Consider a system of \(N\) nonlinear equations,
\begin{equation}
    F_i\bigl(x_0,x_1,\dots,x_{N-1}\bigr) = 0,
    \quad i = 0,1,\dots,N-1.
\end{equation}
We write the vector of unknowns as \(\mathbf{x}\) and the vector of functions as \(\mathbf{F}(\mathbf{x})\).  In the neighborhood of \(\mathbf{x}\), a first order Taylor expansion gives
\begin{equation}
    \mathbf{F}(\mathbf{x} + \delta\mathbf{x})
      = \mathbf{F}(\mathbf{x})
      + \sum_{j=0}^{N-1} \frac{\partial F_i}{\partial x_j}\,\delta x_j
      + \mathcal{O}(\|\delta\mathbf{x}\|^2).
\end{equation}
The partial derivatives define the Jacobian matrix,
\begin{equation}
    J_{ij} = \frac{\partial F_i}{\partial x_j},
\end{equation}
so that
\begin{equation}
    \mathbf{F}(\mathbf{x} + \delta\mathbf{x})
      = \mathbf{F}(\mathbf{x}) + \mathbf{J}\,\delta\mathbf{x}
      + \mathcal{O}(\|\delta\mathbf{x}\|^2).
\end{equation}
Setting \(\mathbf{F}(\mathbf{x} + \delta\mathbf{x}) = \mathbf{0}\) and neglecting higher order terms yields the linear system
\begin{equation}
    \mathbf{J}\,\delta\mathbf{x} = -\,\mathbf{F}(\mathbf{x}),
\end{equation}
which can be solved for \(\delta\mathbf{x}\) by LU decomposition.  Updating the solution,
\begin{equation}
    \mathbf{x}_{\rm new} = \mathbf{x}_{\rm init} + \delta\mathbf{x},
\end{equation}
and iterating this process yields successive approximations until the desired accuracy is achieved.  

\subsection{Laguerre method for root finding}
Consider a polynomial of degree \(n\) defined by
\begin{align}
    P_n(x) &= (x - x_0)(x - x_1)\cdots(x - x_{n-1}),
    \\[4pt]
    \log\bigl|P_n(x)\bigr|
    &= \log\bigl|x - x_0\bigr| + \log\bigl|x - x_1\bigr| + \cdots + \log\bigl|x - x_{n-1}\bigr|.
\end{align}
In the Laguerre method we introduce the functions
\begin{align}
    G(x) &= \frac{{\rm d}}{{\rm d}x}\log\bigl|P_n(x)\bigr|
           = \sum_{i=0}^{n-1} \frac{1}{x - x_i}
           = \frac{P_n'(x)}{P_n(x)}, \\[4pt]
    H(x) &= -\frac{{\rm d}^2}{{\rm d}x^2}\log\bigl|P_n(x)\bigr|
           = \sum_{i=0}^{n-1} \frac{1}{(x - x_i)^2}
           = \left[\frac{P_n'(x)}{P_n(x)}\right]^2 - \frac{P_n''(x)}{P_n(x)}.
\end{align}
Assume that one root, \(x_0\), lies at distance \(a\) from the current estimate \(x\), while all other roots lie at the same 
distance \(b\).  That is,
\[
x - x_0 = a,
\quad
x - x_i = b
\quad (i\neq 0).
\]
Substituting these into the definitions of \(G(x)\) and \(H(x)\) gives
\begin{align}
    G(x) &= \frac{1}{a} + \frac{n-1}{b},
    \\[4pt]
    H(x) &= \frac{1}{a^2} + \frac{n-1}{b^2}.
\end{align}
Solving for \(a\) yields
\begin{equation}
    a = \frac{n}{\,G(x)\pm\sqrt{(n-1)\bigl[n\,H(x)-G^2(x)\bigr]}\,},
\end{equation}
where the sign is chosen to minimize \(|a|\).  Starting from an initial estimate \(x\), one computes \(a\) and updates
\[
x \longleftarrow x - a,
\]
iterating until \(|a|\) is below a chosen tolerance, at which point \(x\approx x_0\).

Polynomial deflation consists of factoring
\[
P_n(x) = (x - r)\,Q_{n-1}(x),
\]
where \(r\) is a found root and \(Q_{n-1}(x)\) is a polynomial of degree \(n-1\).  One then applies the Laguerre method to 
\(Q_{n-1}(x)\) to locate another root of \(P_n(x)\), and repeats until all \(n\) roots are determined.

\subsection{Code structure}
\label{code}
The purpose of the program is to compute the energy of Cooper pairs in the Richardson model using an eigenvalue-based method. The 
system structure (number of Cooper pairs, energy levels, and degeneracies) is read from input \texttt{.txt} files. The source 
code and a guide for preparing the input files are available on GitHub:
\begin{quote}
[\href{https://github.com/GrzegorzBiskowski/Richardson-solver}{https://github.com/GrzegorzBiskowski/Richardson-solver}]. 
\end{quote}
After the computation, the program saves the energy as 
a function of the coupling constant $g$ in both \texttt{.txt} and \texttt{.csv} formats.

\paragraph{General workflow.}
After loading the data, the program initializes the starting value of $g$ and begins the first iteration of a \texttt{do-while} 
loop. It first solves a system of nonlinear equations using the Newton--Raphson method, where the associated linear system is 
handled via LU decomposition. The resulting $\Lambda$ values are then used to build another linear system for the coefficients of 
the polynomials $P(z)$.

Ultimately, the Cooper pair energies are computed using the Laguerre method with polynomial deflation and written to a file. Each 
iteration concludes by incrementing $g$ by a small step. The loop continues until a target value of $g$ is reached. In this 
study, the final value used was $g = 1.5$.

\paragraph{Code architecture.}
The program consists of three main files: \texttt{main.cpp}, \texttt{solver.cpp}, and \texttt{solver.h}. The file 
\texttt{solver.h} contains class declarations, while \texttt{solver.cpp} defines the corresponding methods.

Three main classes were developed:
\begin{itemize}
  \item \texttt{cSqMatrix}
  \item \texttt{cEquations}
  \item \texttt{cPolynomials}
\end{itemize}

\textbf{\texttt{cSqMatrix}} stores square matrices and is responsible for solving linear systems. The right-hand side vector is 
passed via the constructor, which immediately invokes LU decomposition. The solution then replaces the original input vector.

\textbf{\texttt{cEquations}} solves the nonlinear system for $\Lambda$ values. It includes the method \texttt{function\_i}, which 
returns the left-hand side of the equation for a given index $i$ and vector of values, as well as \texttt{newton\_raphson}, which 
updates the \textit{initial guess} with the solution obtained via the Newton--Raphson method.

\textbf{\texttt{cPolynomials}} takes the energy vector of elements $E_j$ (from the previous iteration, or equal to $\epsilon_j$ 
in the first iteration), the value of $g$, and the previously computed $\Lambda$ values. It builds a matrix of type 
\texttt{cSqMatrix} for the polynomial coefficients, which are then calculated using LU decomposition. The Cooper pair energies 
are obtained by finding the roots using the Laguerre method, implemented in the \texttt{root\_finder} function. This class also 
provides helper methods: \texttt{value}, \texttt{first\_der}, and \texttt{second\_der}, which compute the polynomial value, first 
derivative, and second derivative, respectively, for a given argument (the current guess in the Laguerre method).

\paragraph{Libraries and tools.}
The following C++ standard libraries are used:
\begin{itemize}
  \item \texttt{iostream}
  \item \texttt{fstream}
  \item \texttt{iomanip}
  \item \texttt{cmath}
  \item \texttt{complex}
\end{itemize}

The first three handle file I/O and console output. In particular, \texttt{cout} is used to print the current $g$ value to the 
console for monitoring purposes. The \texttt{complex} library is essential for the Laguerre method, which must compute both real 
and complex roots of $P(z)$. Consequently, all program variables are declared as \texttt{std::complex<double>}, including the $
\epsilon$ vector. This also reflects the convention used for energy splitting in the complex plane given by~(\ref{deg}).

\paragraph{Development environment and stability.}
The code was written using the Code::Blocks 20.03 IDE and compiled using \texttt{g++}. The program performs with high numerical 
stability and accuracy for systems with up to 20 Cooper pairs. For larger systems, small discontinuities may occasionally appear 
in the $\Lambda$ solutions, which can influence the precision of the resulting energies. However, this opens the door to future 
improvements, such as fine-tuning numerical parameters or incorporating adaptive stabilization techniques to extend the solver's 
robustness to even larger systems.

It should be noted that all algorithms were deliberately implemented in their most basic form, without incorporating advanced 
numerical stabilization techniques. This simplicity constitutes a practical advantage, as it enables both straightforward use and 
further development of the code without requiring specialized programming or numerical analysis expertise.

\section{Example calculations and applications}
\label{Ch6}
\subsection{Non--degenerate solutions}
The methods described in the previous sections were used to compute the energy spectrum of the Cooper pairs, \(E_J\), as a 
function of the coupling constant \(g\) for various models.\footnote{In this subsection, all energies are given per single 
electron; consequently, at $g=0$ each Cooper pair has twice the listed energy.}

We first consider the one-dimensional quantum harmonic oscillator with \(L=12\) levels,
\[
\epsilon_j = j + \tfrac{1}{2},
\quad j = 0,1,\dots,11,
\]
occupied by \(M=6\) Cooper pairs filling the lowest available levels.
\begin{figure}[h]
    \centering
    \begin{minipage}{.45\textwidth}
        \centering
        \scalebox{0.55}{\input{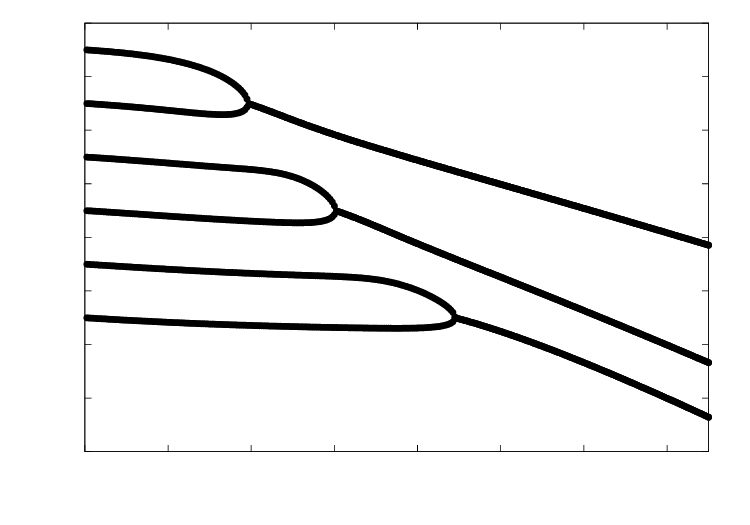}}
        \caption{Real Cooper pair energies for the picket-fence model with $L=12$ levels and $M=6$ pairs.}
        \label{fig:picket_fence_re}
    \end{minipage}\text{\qquad}%
    \begin{minipage}{.45\textwidth}
        \centering
        \scalebox{0.55}{\input{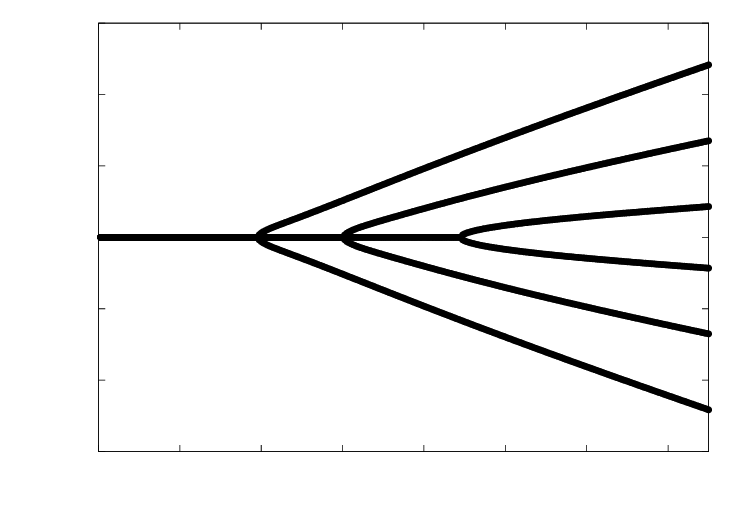}}
        \caption{Imaginary Cooper pair energies for the picket-fence model with $L=12$ levels and $M=6$ pairs.}
        \label{fig:picket_fence_im} 
    \end{minipage}
\end{figure}
Figures \ref{fig:picket_fence_re} and \ref{fig:picket_fence_im} display the real and imaginary parts of the Cooper pair energies, 
respectively. The real parts begin to coalesce at different values of \(g\), depending on the initial energy levels; these 
characteristic arcs also appear in subsequent examples.

According to the Richardson equations, a singularity would occur when \(E_{J_\nu} = E_{J_\mu}\) for \(\mu\neq\nu\).  However, by 
allowing the energies to extend into the complex plane, the imaginary parts branch out precisely at the points where two real 
parts merge.  Figure \ref{fig:picket_fence_im} shows that this branching is symmetric, thereby preserving the reality of the 
total energy.  
\begin{figure}[h]
    \centering
    \begin{minipage}{.45\textwidth}
        \centering
        \scalebox{0.55}{\input{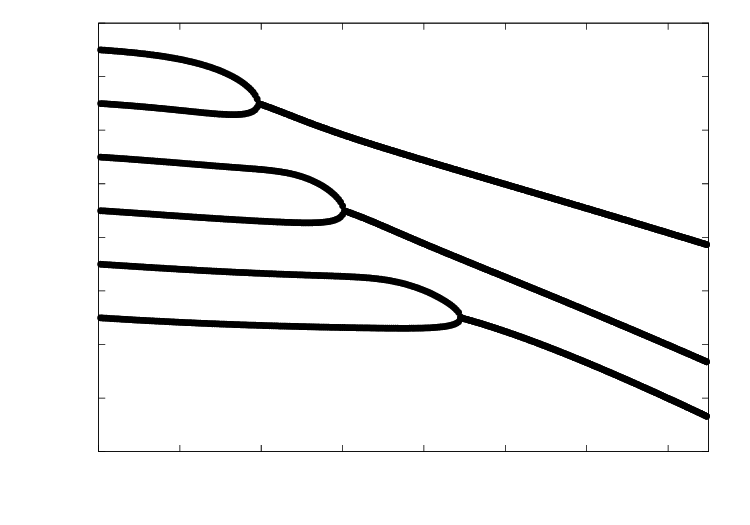}}
        \caption{Real Cooper pair energies for the picket-fence model with $L=12$ levels and $M=6$ pairs.}
        \label{fig:10g_picket_fence_re}
    \end{minipage}\text{\qquad}%
    \begin{minipage}{.45\textwidth}
        \centering
        \scalebox{0.55}{\input{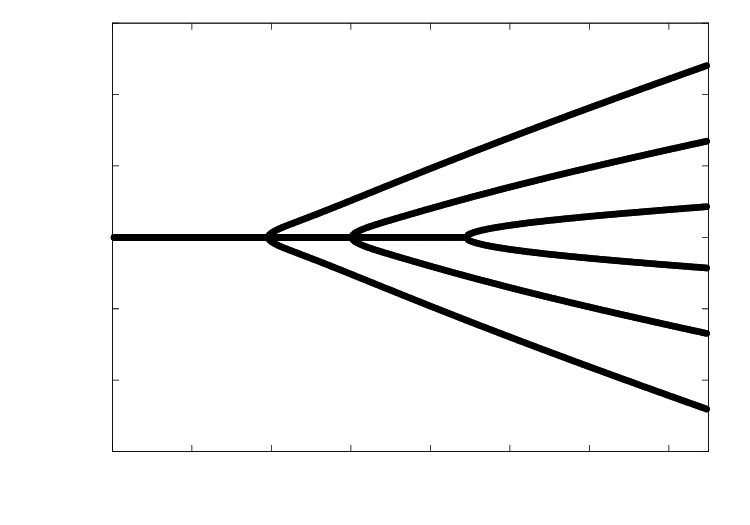}}
        \caption{Imaginary Cooper pair energies for the picket-fence model with $L=12$ levels and $M=6$ pairs.}
        \label{fig:10g_picket_fence_im} 
    \end{minipage}
\end{figure}

In the next example, the single particle energies are given by 
\[
\epsilon_j = 10\times\Bigl(j + \tfrac12\Bigr),
\quad j = 0,1,\dots,11,
\]
and the coupling constant \(g\) is scaled by a factor of 10 in Figures \ref{fig:10g_picket_fence_re} and 
\ref{fig:10g_picket_fence_im}.  Despite multiplying both \(g\) and \(\epsilon_j\) by 10, the resulting spectra are identical.  
\begin{figure}[h]
    \centering
    \begin{minipage}{.45\textwidth}
        \centering
        \scalebox{0.55}{\input{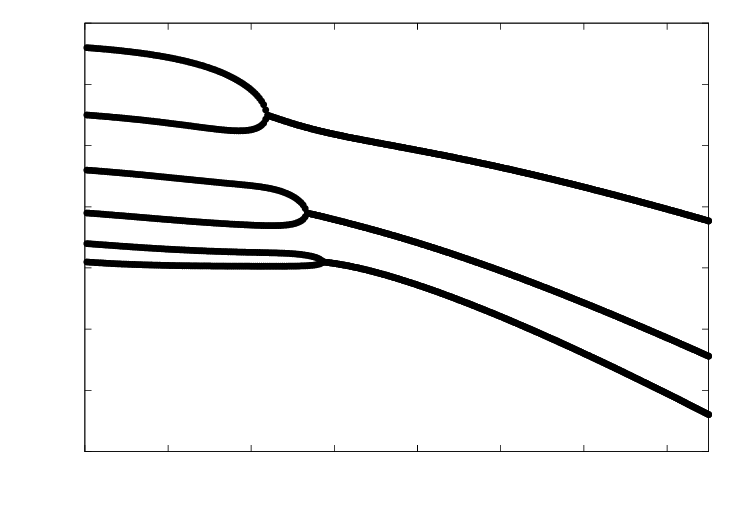}}
        \caption{Real Cooper pair energies for the 1D infinite potential well with $L=12$ levels and $M=6$ pairs.}
        \label{fig:square_well_re}
    \end{minipage}\text{\qquad}%
    \begin{minipage}{.45\textwidth}
        \centering
        \scalebox{0.55}{\input{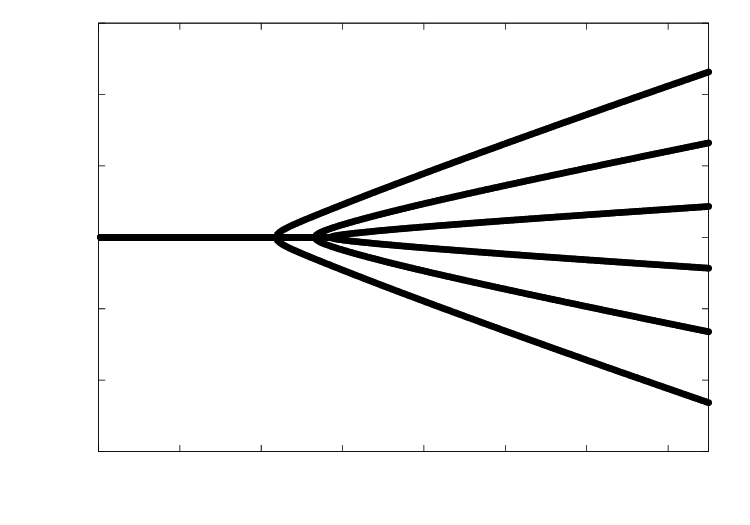}}
        \caption{Imaginary Cooper pair energies for the 1D infinite potential well with $L=12$ levels and $M=6$ pairs.}
        \label{fig:square_well_im} 
    \end{minipage}
\end{figure}
Figures \ref{fig:square_well_re} and \ref{fig:square_well_im} display the real and imaginary parts of the Cooper pair spectrum, 
respectively, for \(M=6\) pairs occupying \(L=12\) levels with single particle energies \(\epsilon_j\propto j^2\), corresponding 
to the infinite potential well model. 

Notice that the lowest lying levels merge at smaller values of \(g\), reflecting the reduced initial energy 
spacing between these levels.

\begin{figure}[h]
    \centering
    \begin{minipage}{.45\textwidth}
        \centering
        \scalebox{0.55}{\input{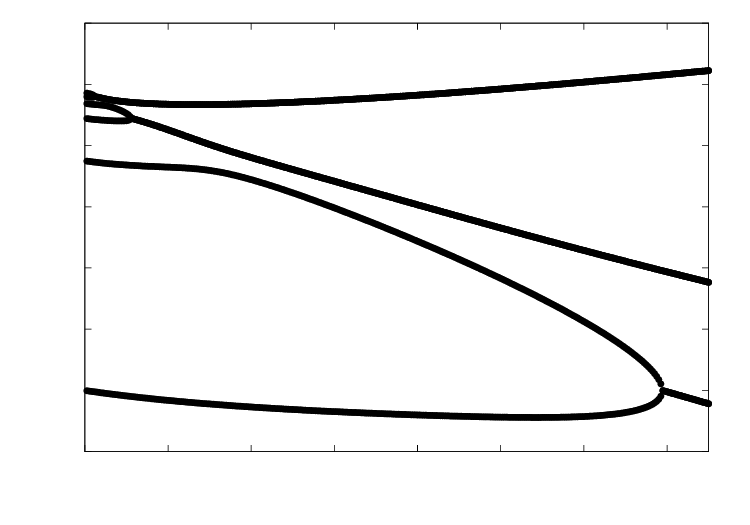}}
        \caption{Real Cooper pair energies for the $\epsilon_j\propto -j^{-2}$ model with $L=6$ levels and $M=6$ pairs.}
        \label{fig:inv_square_re}
    \end{minipage}\text{\qquad}%
    \begin{minipage}{.45\textwidth}
        \centering
        \scalebox{0.55}{\input{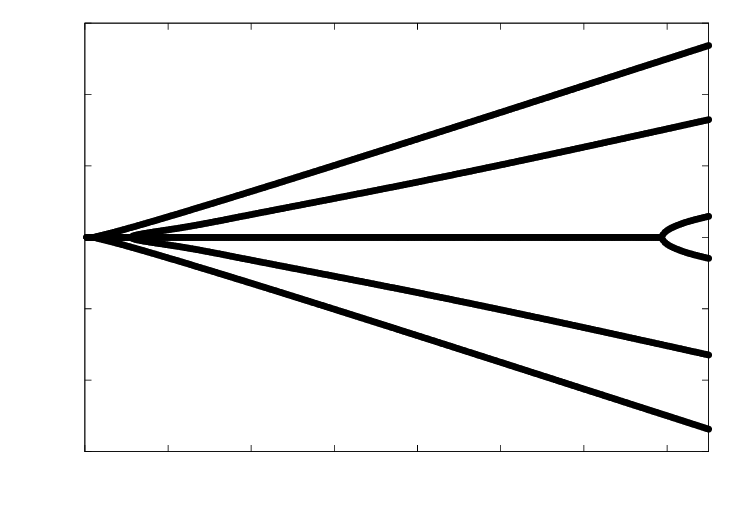}}
        \caption{Imaginary Cooper pair energies for the $\epsilon_j\propto -j^{-2}$ model with $L=6$ levels and $M=6$ pairs.}
        \label{fig:inv_square_im} 
    \end{minipage}
\end{figure}

In the next example, we choose single particle energies to follow a hydrogen-like spectrum,
\[
\epsilon_j \propto -\frac{1}{j^2},
\quad j = 1,2,\dots,
\]
and the resulting real and imaginary parts of the Cooper pair energies are shown in Figures~\ref{fig:inv_square_re} 
and~\ref{fig:inv_square_im}, respectively.
Figures \ref{fig:inv_square_re} and \ref{fig:inv_square_im} display the energies of \(M=6\) Cooper pairs distributed among the 
\(L=6\) highest single particle levels.  The spacing between the two highest levels is extremely small compared to that between 
the two lowest levels, so the former merge almost immediately as \(g\) increases, whereas the latter coalesce only around 
\(g\approx 1.4\).

\subsection{Degenerate solutions}
\begin{figure}[h]
    \centering
    \begin{minipage}{.45\textwidth}
        \centering
        \scalebox{0.55}{\input{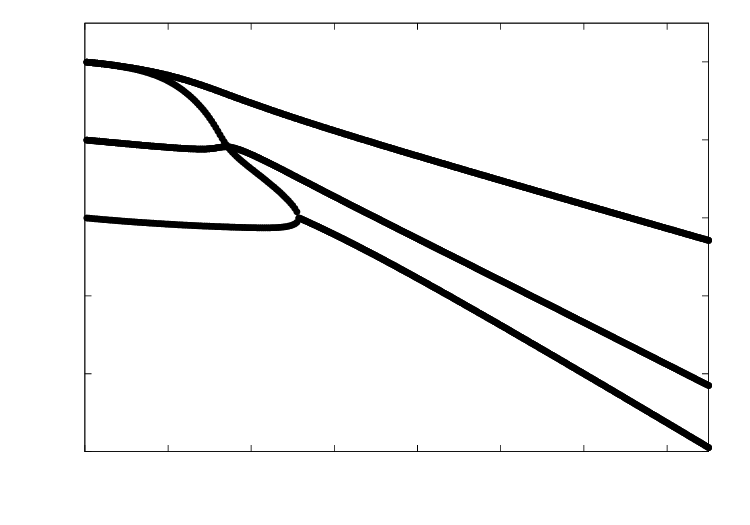}}
        \caption{Real Cooper pair energies for the 2D harmonic oscillator with $L=4$ levels and $M=6$ pairs.}
        \label{fig:2D_oscillator_re}
    \end{minipage}\text{\qquad}%
    \begin{minipage}{.45\textwidth}
        \centering
        \scalebox{0.55}{\input{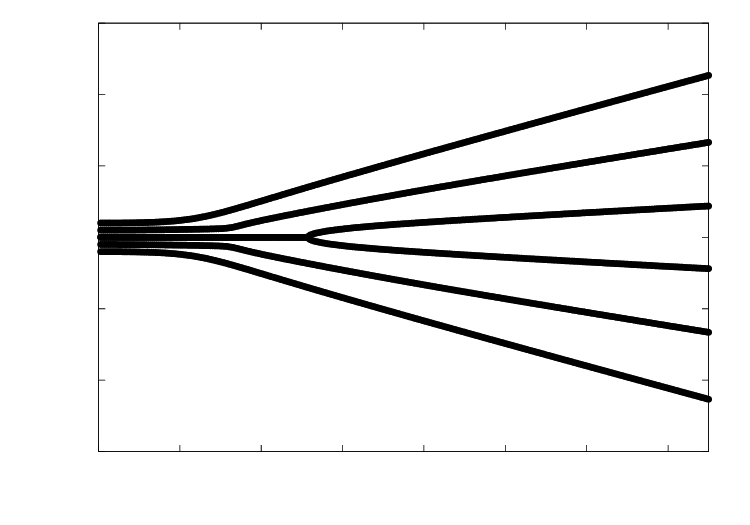}}
        \caption{Imaginary pair energies for the 2D harmonic oscillator with $L=4$ levels and $M=6$ pairs.}
        \label{fig:2D_oscillator_im} 
    \end{minipage}
\end{figure}

Similar calculations were carried out for the degenerate case using the two-dimensional quantum harmonic oscillator.  Here the 
single particle energies scale as 
\[
\epsilon_j = 2j,
\quad
j = 0,1,2,\ldots,
\]
with degeneracies 
\[
D_j = j.
\]
Figures \ref{fig:2D_oscillator_re} and \ref{fig:2D_oscillator_im} show the solutions of the Richardson equations for \(M=6\) 
Cooper pairs distributed over \(L=4\) levels, where the first three levels are fully occupied and the highest level is empty.
Due to the method chosen for handling degeneracies, the imaginary parts of the energies do not start at zero when \(g=0\), as 
they do in the non-degenerate cases. Here, the initial energies are set according to equation \eqref{deg}, 
which modifies the initial values of \(\operatorname{Im}(E_J)\). Nevertheless, the branching behaviour of 
\(\operatorname{Im}(E_J)\) as \(g\) increases closely resembles that seen in the non-degenerate solutions.

For the real parts, we observe that the energies do not fully coalesce as they did in the one-dimensional harmonic oscillator.  
After the first merging at \(g \approx 0.3\), one branch immediately re-splits: one sub-branch merges with the ground state 
energy, while the other remains separate.

\begin{figure}[h]
    \centering
    \begin{minipage}{.45\textwidth}
        \centering
        \scalebox{0.55}{\input{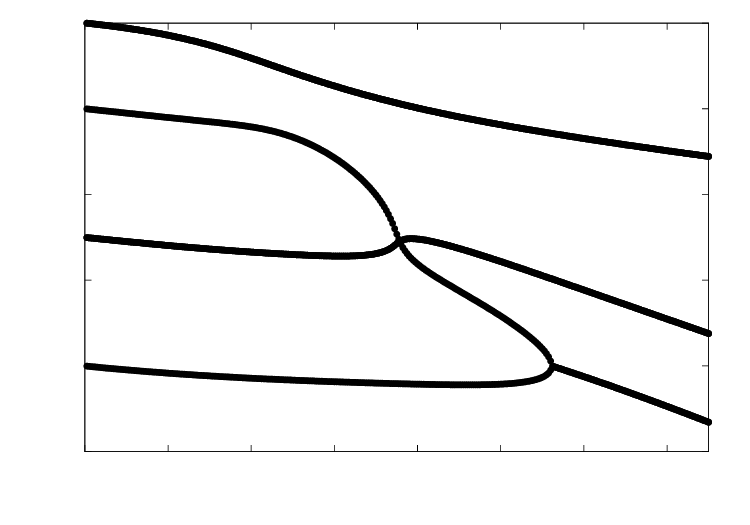}}
        \caption{Real Cooper pair energies for the 2D infinite potential square well with $L=5$ levels and $M=6$ pairs.}
        \label{fig:2D_well_re}
    \end{minipage}\text{\qquad}%
    \begin{minipage}{.45\textwidth}
        \centering
        \scalebox{0.55}{\input{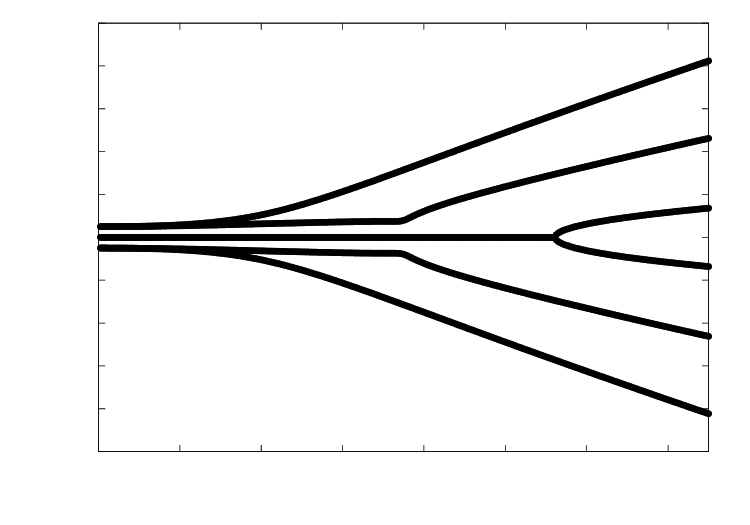}}
        \caption{Imaginary Cooper pair energies for the 2D infinite potential square well with $L=5$ levels and $M=6$ pairs.}
        \label{fig:2D_well_im} 
    \end{minipage}
\end{figure}

We then solved the Richardson equations for the two-dimensional infinite square well, with \(M=6\) Cooper pairs occupying the 
lowest four of the \(L=5\) available levels.  The resulting real and imaginary spectra are presented in Figures 
\ref{fig:2D_well_re} and \ref{fig:2D_well_im}, respectively.
This example was selected because the level degeneracies follow no simple pattern and the energies do not lie on an elementary 
series.  Nevertheless, the behaviour closely resembles that of the two-dimensional quantum harmonic oscillator: pairs of 
solutions coalesce and then immediately branch, with one branch subsequently merging into the lowest energy state. The imaginary 
part of the spectrum exhibits a pronounced cusp near \(g\approx0.8\), which arises from the initial imaginary separations of the 
levels; detailed numerical tests confirm that the magnitude of this initial separation does not affect the real spectrum.

It is also noteworthy that, in both the two-dimensional harmonic oscillator and the two-dimensional infinite square well, certain 
degenerate levels remain inert and do not participate in any merging or branching.

\begin{figure}[h]
    \centering
    \begin{minipage}{.45\textwidth}
        \centering
        \scalebox{0.55}{\input{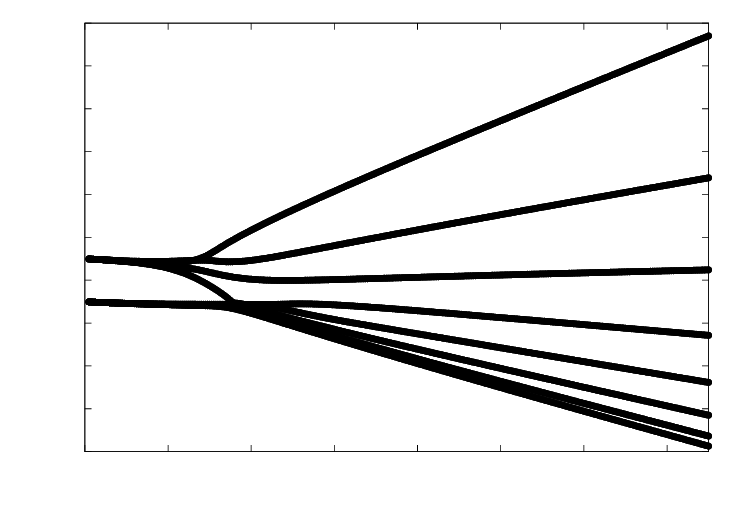}}
        \caption{Real Cooper pair energies for the two-level model with $M=16$ pairs.}
        \label{fig:two_levels_re}
    \end{minipage}\text{\qquad}%
    \begin{minipage}{.45\textwidth}
        \centering
        \scalebox{0.55}{\input{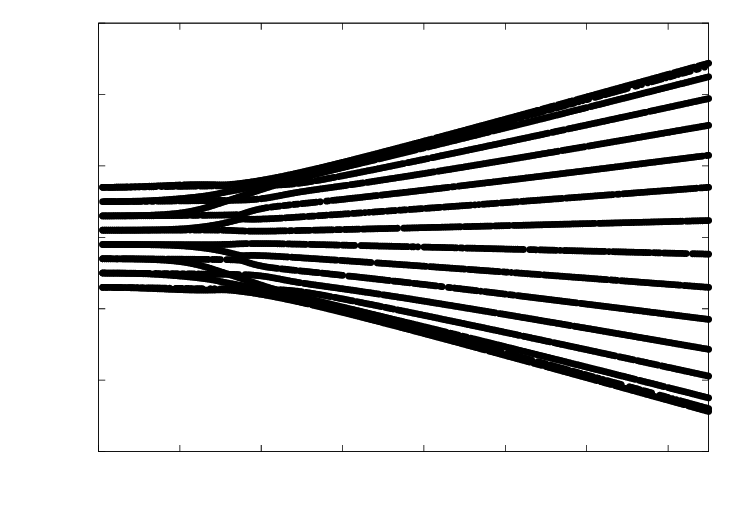}}
        \caption{Imaginary Cooper pair energies for the two-level model with $M=16$ pairs.}
        \label{fig:two_levels_im} 
    \end{minipage}
\end{figure}

Finally, we consider a two-level model with \(M=16\) Cooper pairs equally distributed between the levels \(\epsilon=\pm1\).  The 
resulting real and imaginary parts of the pair energies are shown in Figures \ref{fig:two_levels_re} and \ref{fig:two_levels_im}, 
respectively.
This model does not exhibit the characteristic merging observed previously.  Highly degenerate systems tend to favour branching 
of energy levels, whereas non-degenerate systems promote the merging of the closest levels (in terms of energy gap).  Here, each 
level has degeneracy \(D_J=8\), so the energies branch out and approach a distribution resembling a down-shifted \(j^2\) 
spectrum.  Notably, at \(g=1.5\) the sixteen Cooper pair energies collapse into only eight distinct values, indicating that 
branching stops at pairs of levels rather than proceeding fully. 

\subsection{Thermal pairing}
Thermodynamic effects can be incorporated into the Richardson model by statistically averaging exact many-body eigenenergies 
over a thermal ensemble. Since the total number of fermions $N$ is fixed, the canonical ensemble is the natural framework: it 
preserves particle number while allowing the system's energy to fluctuate according to temperature.

A central observable in this context is the {\it temperature-dependent pairing energy}, defined as the thermal average energy 
difference between the interacting ($g\neq 0$) and non-interacting ($g=0$) systems (cf.~\cite{Ciechan}):
\begin{equation}
E_p(g,T)=\langle E(g)\rangle-\langle E(g=0)\rangle,
\end{equation}
where the thermal average $\langle E\rangle$ is given by
\begin{equation}
\langle E\rangle=\frac{1}{Z}\sum_\alpha D_\alpha E_\alpha \exp\left(-\frac{E_\alpha}{kT}\right),
\end{equation}
and the canonical partition function is
\begin{equation}
Z=\sum_\alpha D_\alpha \exp\left(-\frac{E_\alpha}{kT}\right).
\end{equation}
Here, $E_\alpha$ are the exact eigenenergies of the system, $D_\alpha$ their degeneracies, and $k$ is the Boltzmann constant.

The temperature dependence of the internal energy $\langle E \rangle$ further allows for direct computation of thermodynamic 
quantities. In particular, the {\it heat capacity} at fixed particle number is obtained via
\begin{equation}
C_V(T) = \left.\frac{\partial \langle E \rangle}{\partial T} \right|_{N,V}.
\end{equation}
This framework enables finite-temperature studies of pairing phenomena in finite-size systems, without resorting to mean-field or 
grand canonical approximations. It also provides a concrete benchmark for testing extensions of the Bethe ansatz to thermodynamic 
regimes.

In the case of the two-level Richardson model, all Cooper pair eigenenergies (Bethe rapidities) and their associated degeneracies 
can be determined explicitly. For instance, Ciechan and Wysoki\'{n}ski~\cite{Ciechan} computed 
the complete spectrum for $M=10$ pairs and used direct state summation to evaluate the specific heat and pairing energy as 
functions of $T$. 
Using our solver, we can readily reproduce these results and extend them to larger system sizes---namely, systems with more than 
two energy levels ($L>2$) and over ten pairs ($M>10$)---as well as to finer temperature grids. These results, among others, will 
be presented in a forthcoming paper.

\subsection{Yang--Yang function and classical conformal blocks}
\label{EYY}
As an example of the correspondence between the classical limit of conformal blocks and quantum integrable systems of the 
Richardson--Gaudin type, let us consider the Coulomb gas representation of a spherical Virasoro four-point block:
\begin{align*}
\mathcal{Z}(\,\cdot\,|{\bf z}_f)
&=
\left\langle 
  :{\rm e}^{\hat\alpha_{1}\phi(0)}:\,
  :{\rm e}^{\hat\alpha_{2}\phi(x)}:\,
  :{\rm e}^{\hat\alpha_{3}\phi(1)}:\,
  :{\rm e}^{\hat\alpha_{4}\phi(\infty)}:\,
  \biggl[\!\int_{0}^{x}\!{\rm e}^{b\phi(E)}\,{\rm d}E\biggr]^{M_1}
  \biggl[\!\int_{0}^{1}\!{\rm e}^{b\phi(E)}\,{\rm d}E\biggr]^{M_2}
\right\rangle
\\[6pt]
&=
x^{\frac{\alpha_1\alpha_2}{2\beta}}
(1-x)^{\frac{\alpha_2\alpha_3}{2\beta}}\times
\\[6pt]
&\times
\prod_{\mu=1}^{M_1}\!\int_{0}^{x}\!{\rm d}E_\mu
\prod_{\mu=M_{1}+1}^{M_{1}+M_{2}}\!\int_{0}^{1}\!{\rm d}E_\mu
\prod_{\mu<\nu}\bigl(E_\nu - E_\mu\bigr)^{2\beta}
\;\prod_{\mu}
E_{\mu}^{\alpha_1}
\,(E_\mu-x)^{\alpha_2}
\,(E_\mu-1)^{\alpha_3},
\end{align*}
where ${\bf z}_f=(0,x,1,\infty)$. It was not clear for a long time how to choose integration contours to get 
an integral representation of the four-point block
consistent with historically first BPZ power series representation \cite{BPZ}:
\begin{eqnarray*}
	\label{block}
	{\cal F}\!\left(\Delta_4,\ldots,\Delta_1,\Delta,c\,|\,x\,\right)
	&=&
	x^{\Delta-\Delta_{2}-\Delta_{1}}\left( 1 +
	\sum_{n=1}^\infty x^{n}
	{\cal F}_{n}\!\left(\Delta_4,\ldots,\Delta_1,\Delta,c\right)\right),
	\\
	{\cal F}_{n}\!\left(\Delta_4,\ldots,\Delta_1,\Delta,c\right)&=&
	\sum\limits_{n=|I|=|J|}
	\left\langle\,\Delta_4\,| V_{\Delta_3}(1) |\,\Delta_{I}^{n}\,\right\rangle
	\Big[G^{(n)}_{c,\Delta}\Big]^{IJ}
	\left\langle\,\Delta_{J}^{n}\,| V_{\Delta_2}(1) |\,\Delta_1\,\right\rangle.
\end{eqnarray*}
Here, \(\Big[G^{(n)}_{c,\Delta}\Big]^{IJ}\) denotes the inverse of the Gram matrix
\[
\Big[G^{(n)}_{c,\Delta}\Big]_{IJ}
\;=\;
\bigl\langle \Delta^n_I\!\bigm|\!\Delta^n_J\bigr\rangle,
\]
where 
$$
\ket{\Delta^{n}_{I}}=L_{-I}\ket{\Delta} 
= L_{-i_{k}}\ldots L_{-i_{2}}L_{-i_{1}}\ket{\Delta} 
$$
is the standard basis of the subspace
\[
{\cal V}_{(n,\Delta)}\;\subset\;\bigoplus_{n=0}^\infty\mathcal{V}_{(n,\Delta)}
\]
of the Verma module.  Basis vectors are labeled by partitions
$I = (i_k \ge \cdots \ge i_1 \ge 1)$,
whose total length is \(\lvert I\rvert := i_1 + \cdots + i_k = n\). The symbols
\(V_{\Delta_j}(z)\) denote Virasoro chiral vertex operators,
$$
V_{\Delta_j}(z)=V\left(|\,\Delta_j\,\rangle | z\right)=
V_{\Delta_k,\Delta_i}^{\;\Delta_j}\left(|\,\Delta_j\,\rangle | z\right):
{\cal V}_{\Delta_i}\longrightarrow {\cal V}_{\Delta_k}
$$
that act between Verma modules and are defined by
(i) commutation relations with Virasoro generators,
$$
\left[L_n , V_{\Delta}(z)\right] = z^{n}\left(z
\frac{{\rm d}}{{\rm d}z} + (n+1)\Delta\right)V_{\Delta}(z);
$$ 
(ii) normalization 
$\left\langle\,\Delta_3\,| V_{\Delta_2}(z) |\,\Delta_{1}\,\right\rangle=z^{\Delta_3-\Delta_2-\Delta_1}$.

Mironov, Morozov and Shakirov showed that ${\cal Z}(\,\cdot\,|{\bf z}_f)$ 
precisely reproduces the BPZ four-point block expansion \cite{MMS}. This result holds assuming that 
\begin{eqnarray*}
\Delta_i &=& \frac{\alpha_i\left(\alpha_i+2-2\beta\right)}{4\beta},\quad i=1,2,3,\\
\Delta_4 &=& \frac{\left(2\beta\left(M_1+M_2\right)+\alpha_1+\alpha_2+\alpha_3\right)
\left(2\beta\left(M_1+M_2\right)+\alpha_1+\alpha_2+\alpha_3+2-2\beta\right)}{4\beta},\\
\Delta &=& \frac{\left(2\beta M_1+\alpha_1+\alpha_2\right)
\left(2\beta M_1+\alpha_1+\alpha_2+2-2\beta\right)}{4\beta},\\
c &=& 1-6\left(\sqrt{\beta}-\frac{1}{\sqrt{\beta}}\right)^2,
\quad\beta=b^2, \quad\alpha_i=2b\hat\alpha_i.
\end{eqnarray*}
Thus, there are two ways to compute the $b\to\infty$
asymptotic of ${\cal Z}(\,\cdot\,|{\bf z}_f)$. 
On the one hand, it corresponds to the saddle-point approximation of the integral.
On the other hand, it coincides with the classical (large-$c$) limit of the BPZ four-point conformal block,
\[
\mathcal{F}(\Delta_i,\Delta,c \!\mid\! x)
\;\xrightarrow[b\to\infty]{}\;
\exp\!\Big[b^2 f(\delta_i,\delta\!\mid\! x)\Big],
\]
where we define
\[
\delta_i \;=\;\lim_{b\to\infty}\frac{\Delta_i}{b^2},
\qquad
\delta \;=\;\lim_{b\to\infty}\frac{\Delta}{b^2}.
\]
Equivalently,
\[
f(\delta_i,\delta\!\mid\! x)
=\lim_{b\to\infty}\frac{1}{b^2}\log\mathcal{F}(\Delta_i,\Delta,c\!\mid\! x).
\]
In particular, the small-\(x\) expansion of the classical block reads
\[
f(\delta_i,\delta\!\mid\! x)
=(\delta-\delta_1-\delta_2)\log x
\;+\;\frac{(\delta+\delta_2-\delta_1)\,(\delta+\delta_3-\delta_4)}{2\delta}\;x
\;+\;\mathcal{O}(x^2).
\]
We therefore arrive at the following proposition. Set
\begin{eqnarray}
\delta_i&=&\eta_i(\eta_i-1),\quad i=1,2,3, 
\nonumber\\
\delta_4&=&\bigl(M_1+M_2+\eta_1+\eta_2+\eta_3\bigr)\bigl(M_1+M_2+\eta_1+\eta_2+\eta_3-1\bigr),
\nonumber\\
\delta&=&\bigl(M_1+\eta_1+\eta_2\bigr)\bigl(M_1+\eta_1+\eta_2-1\bigr).
\label{cld}
\end{eqnarray}
If the classical conformal weights are chosen as in \eqref{cld}, then the four-point classical block on the sphere can be 
written in closed form as follows:
\begin{eqnarray}\label{4pt}
f(\delta_i,\delta \!\mid\! x)&=&
-\,{\cal W}_{\rm crit}(\mathbf z_f,\mathbf E^{\rm c})
\;-\;\Bigl[S_{M_1}(2\eta_1,2\eta_2)+S_{M_2}\bigl(2(\eta_1+\eta_2+M_1),2\eta_3\bigr)\Bigr]
\nonumber\\
&+&2\eta_1\eta_2\log x +2\eta_2\eta_3\log(1-x),
\end{eqnarray}
where the YY function is
\begin{eqnarray*}
{\cal W}(\mathbf z_f,\mathbf E)
&=& 
-2\sum_{\mu<\nu}\log(E_\nu-E_\mu)
\\
&-&\sum_{\mu=1}^{M_1+M_2}
\Bigl[
2\eta_1\log E_\mu
+2\eta_2\log(E_\mu-x)
+2\eta_3\log(E_\mu-1)
\Bigr].
\end{eqnarray*}
The function \(S_n(a,b)\) is obtained from the saddle-point asymptotics of the Selberg integral (see~\cite{PNPP}).
The critical parameters \(\mathbf E^{\rm c}=\{E_1^{\rm c},\dots,E_M^{\rm c}\}\), 
with \(M=M_1+M_2\), are the solutions of the saddle-point equations:
\begin{equation}\label{Sed1}
\boxed{\;\frac{\partial {\cal W}(\mathbf z_f,\mathbf E)}{\partial E_\mu}=0
\;\;\;\Leftrightarrow\;\;\;
\frac{2\eta_1}{E_\mu}
+\frac{2\eta_2}{E_\mu-x}
+\frac{2\eta_3}{E_\mu-1}
+\sum_{\substack{\nu=1\\\nu\neq\mu}}^M\frac{2}{E_\mu-E_\nu}=0\;}
\end{equation}
where $\mu=1,\dots,M$.
In \cite{PNPP}, relation \eqref{4pt} was verified in the simplest cases---namely $M_1=M_2=1$ and 
$\hat\alpha_1=\hat\alpha_2=\hat\alpha_3=b\eta$---using elementary Mathematica routines. 
A more detailed numerical analysis is required to explore the analytic structure of the classical block.
This example is not unique: a similar correspondence ties WZW and Virasoro torus blocks to the Bethe ansatz solution of the 
elliptic Calogero--Moser model \cite{takemura2000eigenstates,AldayTachikawa,P1}.\footnote{In the case of the torus 
one-point Virasoro block, a rigorous proof establishes the equivalence between its integral representation and its power series 
expansion~\cite{GRSS}. This result was then leveraged within the probabilistic framework to give a fully rigorous 
construction of the torus one-point block and to prove the existence of its classical limit~\cite{DGP}.} 
One of our main motivations in developing a numerical solver for Bethe equations in RG models was to explore these 
connections. We will present the analytical and numerical results on these correspondences in a forthcoming paper.

\section{Conclusions and outlook}
\label{Ch7}
The present work has begun with a review of classical and quantum integrability, together with an introduction to 
the Richardson reduced BCS and Gaudin central spin models.
To ensure the paper's self-sufficiency, every derivation concerning Richardson's solution has been presented in full detail, 
providing a comprehensive reference for readers new to the subject.
Building on this foundation, we have reexamined the well-established correspondence between Richardson--Gaudin models and 
two-dimensional conformal field theory and have identified the appearance of irregular, degenerate Virasoro blocks within this 
framework. 

Our principal achievement is the development of a numerical solver for the Bethe equations in Richardson--Gaudin models.
Our numerical experiments confirm that the implementation yields accurate rapidity trajectories. 
Of particular note is the observation that the rapidity trajectories are invariant with respect to scaling---a property with 
potentially profound practical and physical implications that deserves further investigation. 
A possible direction for future research is to leverage the electrostatic analogy to design novel and more robust numerical 
methods for computing rapidities.\footnote{Indeed, one can interpret the $E_\nu$ as mobile unit charges that 
repel each other via a logarithmic Coulomb interaction, while being attracted to fixed background charges located at the $\epsilon_j$. The variational 
condition $\nabla\mathcal{W}^{\rm R}(\{E_\nu\})=0$ then corresponds to finding the electrostatic equilibrium configuration of 
this charge system. This physical picture naturally suggests alternative numerical strategies. In particular, modern 
gradient-based optimization algorithms---especially when combined with automatic differentiation and appropriate ``soft-core'' 
regularization to avoid numerical singularities when $E_\mu \approx E_\nu$---provide a promising alternative to traditional 
Newton--Raphson or Laguerre solvers. Such methods may offer improved robustness, better convergence behavior, and scalability to 
large systems.}

We have also described an extension of our method to finite temperatures, introducing a scheme for directly computing 
temperature-dependent pairing energies and other thermodynamic observables within the discrete Richardson model.
It would be interesting to reexamine time-dependent extensions of the Richardson--Gaudin models and to explore their potential 
connections with the 2D CFT framework. Such a correspondence appears plausible, especially in view of the formulation in terms of 
the Yang--Yang function (see~e.g.~\cite{BBGMY} and refs.~therein). 

This work adopts a standard analytical-numerical perspective, combining Bethe ansatz techniques with computational tools to 
study Richardson--Gaudin integrable models. However, our broader program envisions three complementary directions for future 
exploration:
\begin{enumerate}
\item {\it Matrix model perspective.}\\
One first observes that, in the thermodynamic limit $L,M\To\infty$ with fixed filling fraction $M/L$, the Bethe rapidities 
$\{E_\nu\}$ condense onto continuous arcs in the complex plane (cf.~\cite{Roman2002}). 
These arcs are described by a spectral density $\rho(E)$ 
supported on a hyperelliptic curve---mirroring the large-$N$ behavior of Hermitian one-matrix model, 
where eigenvalues form cuts on a similar curve via loop (saddle-point) equations \cite{DiFGZ,EKR}.

Moreover, a precise identification follows: the Bethe roots coincide with the zeros of the Baxter polynomial (\ref{Bax}), 
which itself satisfies a second order differential equation. In the semiclassical (large-$N$) limit, this differential 
equation reproduces the spectral curve of the corresponding matrix model.

An interesting problem is to establish this Bethe ansatz \(\leftrightarrow\) matrix model mapping 
rigorously for integrable pairing models and to 
assess its practical utility. Encouragingly, known results for Gaudin-type models (see~\cite{Jurco:2004}) 
suggest this program is within reach.

\item{\it Advanced techniques from 2D CFT.}\\
The appearance of Gaiotto--Witten irregular, degenerate conformal blocks in the structure of the Richardson Yang--Yang 
function opens a path to apply powerful tools from 2D CFT. 
Our primary objective is to exploit the BPZ differential equations satisfied by these blocks to derive---via their classical 
(large-$c$) limit and a subsequent saddle-point analysis---the equations for the critical Richardson Yang--Yang function 
$\mathcal{W}^{\mathrm{R}}_{\mathrm{crit}}$. Moreover, the Gaiotto--Witten irregular blocks featured here have recently 
attracted renewed attention due to their connection with a novel class of integrable systems (see~\cite{HLR,GHLR}). 
It would be highly worthwhile to investigate how these emerging structures relate to---and potentially 
enrich---the Richardson--Gaudin models.

It would also be interesting to examine the RG models from the perspective of the matrix model 
($\beta$-ensemble)/2D CFT correspondence developed in \cite{MS2017}.

\item{\it Connections with 4D gauge theories.}\\
Through the AGT \cite{AGT} and Bethe/gauge \cite{NS:2009} correspondences, 
Richardson--Gaudin systems may also find reinterpretation in terms of four-dimensional \(\mathcal{N}=2\) supersymmetric gauge 
theories. These theories provide a unifying framework for linking matrix 
models, integrable systems, and 2D CFT. For example, the Seiberg--Witten curves---analogs of matrix model spectral curves---and 
their quantum deformations naturally encode the spectra of quantum integrable systems, suggesting a bridge between 
$\mathcal{N}=2$ gauge theories and Richardson--Gaudin models.
\end{enumerate}

\section*{Acknowledgements}
The research presented here has been supported by the Polish National Science Centre under grant no.~2023/49/B/ST2/01371.

\appendix
\section{Derivation of the Poisson bracket of Lax matrices}
\label{A}
We aim to compute the Poisson bracket of Lax matrices:
\[
\{L_1, L_2\} = [r_{12}, L_1] - [r_{21}, L_2],
\]
where \( L = U \Lambda U^{-1} \) is the Lax matrix, with \( \Lambda \) being a diagonal matrix of eigenvalues and \( U \) a 
matrix of eigenvectors.

We begin by evaluating:
\[
\{L_1, L_2\} = \{U_1 \Lambda_1 U_1^{-1}, U_2 \Lambda_2 U_2^{-1}\}.
\]
Applying the Leibniz rule:
\begin{align*}
\{L_1, L_2\} &= \{U_1 \Lambda_1 U_1^{-1}, U_2 \Lambda_2 U_2^{-1}\} \\
&= U_1 \{\Lambda_1 U_1^{-1}, U_2 \Lambda_2 U_2^{-1} \} + \{U_1, U_2 \Lambda_2 U_2^{-1} \} \Lambda_1 U_1^{-1}.
\end{align*}
Expanding all brackets using the Leibniz and chain rules, we obtain a sum of nine terms. To simplify the expression, we define:
\[
k_{12} \equiv \{U_1, U_2\} U_1^{-1} U_2^{-1}, \qquad k_{21} \equiv \{U_2, U_1\} U_2^{-1} U_1^{-1},
\]
\[
q_{12} \equiv U_2 \{U_1, \Lambda_2\} U_2^{-1} U_1^{-1}, \qquad q_{21} \equiv U_1 \{U_2, \Lambda_1\} U_1^{-1} U_2^{-1}.
\]
Note that \( k_{12} = -k_{21} \), while no such relation generally holds between \( q_{12} \) and \( q_{21} \).
We also use the identity:
\[
\{A_1, B_2^{-1}\} = B_2^{-1} \{A_1, B_2\} B_2^{-1},
\]
which follows from \( \{A_1, B_2^{-1} B_2\} = 0 \).
Now we compute each term in the Poisson bracket:
\begin{enumerate}
    \item \( U_1 \Lambda_1 U_2 \Lambda_2 \{U_1^{-1}, U_2^{-1}\} = L_1 L_2 k_{12} \)
    \item \( U_1 \Lambda_1 U_2 \{U_1^{-1}, \Lambda_2\} U_2^{-1} = -L_1 q_{12} \)
    \item \( U_1 \Lambda_1 \{U_1^{-1}, U_2\} \Lambda_2 U_2^{-1} = -L_1 k_{12} L_2 \)
    \item \( U_1 U_2 \Lambda_2 \{\Lambda_1, U_2^{-1}\} U_1^{-1} = L_2 q_{21} \)
    \item \( \{\Lambda_1, \Lambda_2\} = 0 \)
    \item \( U_1 \{\Lambda_1, U_2\} \Lambda_2 U_2^{-1} U_1^{-1} = -q_{21} L_2 \)
    \item \( U_2 \Lambda_2 \{U_1, U_2^{-1}\} \Lambda_1 U_1^{-1} = -L_2 k_{12} L_1 \)
    \item \( U_2 \{U_1, \Lambda_2\} U_2^{-1} \Lambda_1 U_1^{-1} = q_{12} L_1 \)
    \item \( \{U_1, U_2\} \Lambda_2 U_2^{-1} \Lambda_1 U_1^{-1} = k_{12} L_2 L_1 \)
\end{enumerate}
Combining all the terms:
\begin{align*}
\{L_1, L_2\} &= L_1 L_2 k_{12} - L_1 q_{12} - L_1 k_{12} L_2 + L_2 q_{21} - q_{21} L_2 \\
&\quad - L_2 k_{12} L_1 + q_{12} L_1 + k_{12} L_2 L_1.
\end{align*}
Rewriting and grouping:
\begin{align*}
\{L_1, L_2\} &= L_1 [L_2, k_{12}] + [q_{12}, L_1] + [L_2, q_{21}] - [L_2, k_{12}] L_1 \\
&= [L_1, [L_2, k_{12}]] + [q_{12}, L_1] - [q_{21}, L_2].
\end{align*}
By the Jacobi identity:
\[
[[k_{12}, L_2], L_1] + [[L_1, k_{12}], L_2] + [[L_2, L_1], k_{12}] = 0.
\]
Since \( [L_1, L_2] = 0 \), it follows:
\[
[[k_{12}, L_2], L_1] = -[[L_1, k_{12}], L_2] = -[[k_{21}, L_1], L_2].
\]
Thus,
\begin{align*}
\{L_1, L_2\} &= \frac{1}{2} [[k_{12}, L_2], L_1] - \frac{1}{2} [[k_{21}, L_1], L_2] + [q_{12}, L_1] - [q_{21}, L_2] \\
&= \left[ q_{12} + \frac{1}{2}[k_{12}, L_2], L_1 \right] - \left[ q_{21} + \frac{1}{2}[k_{21}, L_1], L_2 \right].
\end{align*}
We define the classical \( r \)--matrix as:
\[
r_{ij} \equiv q_{ij} + \frac{1}{2}[k_{ij}, L_j],
\]
so the Poisson bracket becomes:
\[
\boxed{\{L_1, L_2\} = [r_{12}, L_1] - [r_{21}, L_2].}
\]

\section{Hard--core boson algebra}
\label{AppB1}
\subsection*{Commutator $[b_i,b_j^{\dag}]$}
For the off-diagonal case $i\neq j$, one finds
\begin{align}
\left[b_i, b_j^{\dag}\right] &= b_{i}b_{j}^{\dag} - b_{j}^{\dag}b_i 
= b_{i}b_{j}^{\dag} - c_{j+}^{\dag}c_{j-}^{\dag}c_{i-}c_{i+} 
= b_{i}b_{j}^{\dag} + c_{j+}^{\dag}c_{i-}c_{j-}^{\dag}c_{i+} 
\nonumber\\
&= b_ib_j^{\dag} - c_{i-}c_{j+}^{\dag}c_{j-}^{\dag}c_{i+} 
= b_ib_j^{\dag} + c_{i-}c_{j+}^{\dag}c_{i+}c_{j-}^{\dag} 
= b_ib_j^{\dag} - c_{i-}c_{i+}c_{j+}^{\dag}c_{j-}^{\dag} 
\nonumber\\
&= b_{i}b_{j}^{\dag} - b_{i}b_{j}^{\dag} = 0.
\end{align}
When $i=j$, one obtains
\begin{align}
\left[b_i, b_i^{\dag}\right] &= b_{i}b_{i}^{\dag} - b_{i}^{\dag}b_i 
= b_{i}b_{i}^{\dag} - c_{i+}^{\dag}c_{i-}^{\dag}c_{i-}c_{i+} 
= b_{i}b_{i}^{\dag} - c_{i+}^{\dag}\left(1-c_{i-}c_{i-}^{\dag}\right)c_{i+}\nonumber
\\
&= b_{i}b_{i}^{\dag} - c_{i+}^{\dag}c_{i+} + c_{i+}^{\dag}c_{i-}c_{i-}^{\dag}c_{i+} 
= b_{i}b_{i}^{\dag} - c_{i+}^{\dag}c_{i+} - c_{i-}c_{i+}^{\dag}c_{i-}^{\dag}c_{i+} 
\nonumber\\
&= b_{i}b_{i}^{\dag} - c_{i+}^{\dag}c_{i+} + c_{i-}c_{i+}^{\dag}c_{i+}c_{i-}^{\dag} 
= b_{i}b_{i}^{\dag} - c_{i+}^{\dag}c_{i+} + c_{i-}\left(1-c_{i+}c_{i+}^{\dag}\right)c_{i-}^{\dag} 
\nonumber\\
&= b_{i}b_{i}^{\dag} - c_{i+}^{\dag}c_{i+} + c_{i-}c_{i-}^{\dag} - c_{i-}c_{i+}c_{j+}^{\dag}c_{i-}^{\dag} 
= b_{i}b_{i}^{\dag} - c_{i+}^{\dag}c_{i+} + c_{i-}c_{i-}^{\dag} - b_{i}b_{i}^{\dag} 
\nonumber\\
&= c_{i-}c_{i-}^{\dag} - c_{i+}^{\dag}c_{i+} = 1- c_{i-}^{\dag}c_{-} - c_{i+}^{\dag}c_{i+}. 
\end{align}
Finally, consider the operator
$$
\hat{n}_{i-} + \hat{n}_{i+}
\;=\;
c_{i-}^\dagger c_{i-}\;+\;c_{i+}^\dagger c_{i+}
$$
acting on a given many-body state. Since we restrict our attention to either fully occupied or completely empty levels, each 
number operator $c_{i\sigma}^\dagger c_{i\sigma}$ simply measures the occupancy of the single particle state $\ket{i,\sigma}$. In 
a fully occupied level $\ket{i,-}$ and $\ket{i,+}$, both $\hat{n}_{i-}$ and $\hat{n}_{i+}$ return 1. Conversely, on an empty 
level they each return 0. Thus $\hat{n}_{i-}+\hat{n}_{i+}$ equals 2 for a doubly occupied site and 0 for an unoccupied site.

The operator $b_{i}^{\dag}b_{i}$ refers to the number of Cooper pairs, which occupy certain energy level $\epsilon_i$. This means 
that for fully occupied or non-occupied level, we have
\begin{equation}
c_{i-}^{\dag}c_{i-} + c_{i+}^{\dag}c_{i+} = 2b_{i}^{\dag}b_{i} \;\;\;\Rightarrow\;\;\;\left[b_{i}, b_i^{\dag}\right] 
= 1 - 2b_{i}^{\dag}b_{i}.
\end{equation}
Combining both cases, we obtain
\begin{equation}
\left[b_{i},b_{j}^{\dag}\right] = \delta_{ij}\left(1-2b_i^{\dag}b_{i}\right).
\end{equation}

\subsection*{Commutator $[b_{i}^{\dag}b_i, b_j^{\dag}]$}
First, let us consider the commutator $\left[b_i^{\dag},b_j^{\dag}\right]$:
\begin{align}
\left[b_i^{\dag},b_j^{\dag}\right] &= b_i^{\dag}b_j^{\dag} - b_j^{\dag}b_i^{\dag} 
= b_i^{\dag}b_j^{\dag} - c_{j+}^{\dag}c_{j-}^{\dag}c_{i+}^{\dag}c_{i-}^{\dag} 
= b_i^{\dag}b_j^{\dag} + c_{j+}^{\dag}c_{i+}^{\dag}c_{j-}^{\dag}c_{i-}^{\dag} 
\nonumber\\
&=b_i^{\dag}b_{j}^{\dag} - c_{i+}^{\dag}c_{j+}^{\dag}c_{j-}^{\dag}c_{i-}^{\dag} 
= b_{i}^{\dag}b_{j}^{\dag} + c_{i+}^{\dag}c_{j+}^{\dag}c_{i-}^{\dag}c_{j-}^{\dag} 
= b_{i}^{\dag}b_{j}^{\dag} - c_{i+}^{\dag}c_{i-}^{\dag}c_{j+}^{\dag}c_{j-}^{\dag} 
\nonumber\\
&= b_i^{\dag}b_j^{\dag} - b_{i}^{\dag}b_j^{\dag} = 0 
\;\;\;\Rightarrow\;\;\;
b_{i}^{\dag}b_{j}^{\dag} = b_{j}^{\dag}b_i^{\dag}. 
\end{align}
Using this result, let us first compute the commutator $\bigl[b_i^\dagger b_i,\;b_j^\dagger\bigr]$ in the case $i\neq j$:
\begin{align}
    \left[b_{i}^{\dag}b_i, b_j^{\dag}\right] &= b_i^{\dag}b_ib_j^{\dag} - b_j^{\dag}b_i^{\dag}b_i = b_i^{\dag}b_ib_j^{\dag} - b_i^{\dag}b_j^{\dag}b_i = b_i^{\dag}b_ib_j^{\dag} - b_i^{\dag}b_ib_j^{\dag} = 0.
\end{align}
When \(i=j\), we have
\begin{align}
\left[b_i^{\dag}b_i,b_i^{\dag}\right] &= b_i^{\dag}b_ib_i^{\dag} - b_i^{\dag}b_i^{\dag}b_i 
= b_i^{\dag}\left(1-2b_i^{\dag}b_i + b_i^{\dag}b_i\right) - \left(b_i^{\dag}\right)^2 b_i 
\nonumber\\
&=b_i^{\dag} - b_i^{\dag}b_i^{\dag}b_i = b_i^{\dag} - \left(b_i^{\dag}\right)^2b_i = b_i^{\dag}.
\end{align}
Combining these two results, we get
\begin{equation}
\left[b_i^{\dag}b_i,b_j^{\dag}\right] = \delta_{ij}b_{i}^{\dag}.
\end{equation}

\section{Derivation of the Richardson equations}
\label{AppB2}
Let us consider the eigenvalue problem for the Hamiltonian (\ref{HU}):
\begin{equation}
H_{\cal U}\ket{\Psi_M}_{\cal U} = \mathbb{E}(M)\ket{\Psi_M}_{\cal U}.
\end{equation}
We will use the Bethe ansatz in the form
\begin{equation}
\ket{\Psi_M}_{\cal U} = \prod_{\nu=1}^{M}B_{\nu}^{\dag}\ket{0},
\end{equation}
where
\begin{equation}
B_{\nu}^{\dag} = \sum_{j\in \cal U}\frac{b_j^{\dag}}{2\epsilon_j -E_{\nu}},\text{\qquad} \left[b_j^{\dag}b_j, B_{\nu}^{\dag}
\right] = \frac{b_j^{\dag}}{2\epsilon_j - E_{\nu}}.
\end{equation}
The eigenvalue problem can be rewritten as
\begin{equation}\label{eq: rich-bethe_ansatz_schrodinger}
H_{\cal U}\prod_{\nu=1}^{M}B_{\nu}^{\dag}\ket{0} = \mathbb{E}(M)\prod_{\nu=1}^{M}B_{\nu}^{\dag}\ket{0}.
\end{equation}
Since $H_{\cal U}\ket{0}=0$, we have
\begin{equation}
\left[H_{\cal U},\prod_{\nu=1}^{M}B_{\nu}^{\dag}\right]\ket{0} = H_{\cal U}\prod_{\nu=1}^{M}B_{\nu}^{\dag}\ket{0} - 
\prod_{\nu=1}^{M}B_{\nu}^{\dag}H_{\cal U}\ket{0} = H_{\cal U}\prod_{\nu=1}^{M}B_{\nu}^{\dag}\ket{0}.
\end{equation}
We can also show that the commutator of an operator with a product of operators is generally given by
\begin{equation}
\left[A,\prod_{\nu=1}^{M}B_i\right] = \sum_{\nu=1}^{M}\left(\prod_{\eta=1}^{\nu-1}B_{\eta}\right)\left[A,B_i\right]\left(\prod_{\mu=\nu+1}^{M}B_{\mu}\right).
\end{equation}
Consequently, equation (\ref{eq: rich-bethe_ansatz_schrodinger}) can be rewritten in the form
\begin{equation}\label{eq: rich-hamiltonian_prod}
H_{\cal U}\ket{\Psi_M}_{\cal U} = \sum_{\nu=1}^{M}\left(\prod_{\eta=1}^{\nu-1}B_{\eta}^{\dag}\right)\left[H_{\cal U},B_{\nu}^{\dag}\right]\left(\prod_{\mu=\nu+1}^{M}B_{\mu}^{\dag}\right)\ket{0}.
\end{equation}
To find the commutator $[H_{\cal U},B_{\nu}^{\dag}]$, we introduce the following collective operators:
\begin{equation}
B_0^{\dag} = \sum_{j\in \cal U}b_j^{\dag},\text{\qquad}B_0 = \sum_{j\in \cal U}b_j.
\end{equation}
Accordingly, the Hamiltonian $H_{\cal U}$ can be written as
\begin{align}
H_{\cal U} &= \sum_{j\in \cal U}2\epsilon_jb_j^{\dag}b_j -g\sum_{i,j\in \cal U}b_i^{\dag}b_j 
= \sum_{j\in \cal U}2\epsilon_j b_j^{\dag}b_j - g\sum_{i\in \cal U}b_i^{\dag} \sum_{j\in \cal U}b_j 
\nonumber\\
&= \sum_{j\in \cal U}2\epsilon_jb_j^{\dag}b_j - gB_0^{\dag}B_0.
\end{align}
For simplicity, we first compute the commutators $[B_0,B_{\nu}^{\dag}]$ and $[B_0^{\dag},B_{\nu}^{\dag}]$:
\begin{align}
\left[B_0,B_{\nu}^{\dag}\right] &= B_0B_{\nu}^{\dag} - {\nu}^{\dag}B_0 
= \sum_{i\in \cal U}b_i\sum_{j\in \cal U}\frac{b_j^{\dag}}{2\epsilon_j - E_{\nu}} 
- \sum_{j\in \cal U}\frac{b_j^{\dag}}
{2\epsilon_j - E_{\nu}}\sum_{i\in \cal U}b_i 
\nonumber\\
&= \sum_{i,j\in\cal U}\frac{b_ib_j^{\dag}}{2\epsilon_j - E_{\nu}} 
- \sum_{i,j\in \cal U}\frac{b_j^{\dag}b_i}{2\epsilon_j-E_{\nu}} 
= \sum_{i,j\in \cal U}\frac{b_ib_j^{\dag}-b_j^{\dag}b_i}{2\epsilon_j - E_{\nu}} 
= \sum_{i,j\in \cal U}\frac{\left[b_i,b_j^{\dag}\right]}{2\epsilon_j-E_{\nu}} 
\nonumber\\
&=\sum_{i,j\in \cal U}\frac{\delta_{ij}\left(1-2b_i^{\dag}b_i\right)}{2\epsilon_j - E_{\nu}} 
=\sum_{j\in \cal U}\frac{1-2b_j^{\dag}b_j}{2\epsilon_j - E_{\nu}}, 
\end{align}
and
\begin{align}
\left[B_0^{\dag}, B_{\nu}^{\dag}\right] &= B_0^{\dag}B_{\nu}^{\dag}-B_{\nu}^{\dag}B_0^{\dag} 
= \sum_{i\in \cal U}b_i^{\dag}\sum_{j\in \cal U}\frac{b_j^{\dag}}{2\epsilon_j-E_{\nu}} - \sum_{j\in \cal U}\frac{b_j^{\dag}}{2\epsilon_j-E_{\nu}}\sum_{i\in \cal U}b_i^{\dag} 
\nonumber\\
&=\sum_{i,j \in \cal U}\frac{b_i^{\dag}b_j^{\dag}-b_j^{\dag}b_i^{\dag}}{2\epsilon_j-E_{\nu}} 
= \sum_{i,j\in \cal U}\frac{\left[b_i^{\dag},b_j^{\dag}\right]}{2\epsilon_j-E_{\nu}} 
= 0.
\end{align}
Therefore, the commutator $[H_{\cal U},B_{\nu}^{\dag}]$ is
\begin{align}
\left[H_{\cal U},B_{\nu}^{\dag}\right] 
&= \left[\sum_{i\in \cal U}2\epsilon_ib_i^{\dag}b_i - gB_0^{\dag}B_0, B_{\nu}^{\dag}\right] 
= \left[\sum_{i\in \cal U}2\epsilon_ib_i^{\dag}b_i, B_{\nu}^{\dag}\right] 
- \left[gB_0^{\dag}B_0, B_{\nu}^{\dag}\right] 
\nonumber\\
&= \sum_{i\in \cal U}2\epsilon_i\left[b_i^{\dag}b_i, B_{\nu}^{\dag}\right] 
- gB_0^{\dag}\left[B_0,B_{\nu}^{\dag}\right] 
- g\left[B_0^{\dag},B_{\nu}^{\dag}\right]B_0 
\nonumber\\
&= \sum_{i\in \cal U}2\epsilon_i\left(\frac{b_i^{\dag}}{2\epsilon_i - E_{\nu}}\right) 
- gB_0^{\dag}\sum_{i\in \cal U}\frac{1-2b_i^{\dag}b_i}{2\epsilon_i-E_{\nu}} 
\nonumber\\
&= \sum_{i\in \cal U}\frac{\left(2\epsilon_i - E_{\nu} 
+ E_{\nu}\right)b_i^{\dag}}{2\epsilon_i - E_{\nu}} 
- gB_0^{\dag}\sum_{i\in \cal U}\frac{1-2b_i^{\dag}b_i}{2\epsilon_i - E_{\nu}} 
\nonumber \\
&= E_{\nu}\sum_{i\in \cal U}\frac{b_i^{\dag}}{2\epsilon_i-E_{\nu}} 
+ \sum_{i\in \cal U}\frac{\left(2\epsilon_i - E_{\nu}\right)b_i^{\dag}}{2\epsilon_i - E_{\nu}} 
- gB_0^{\dag}\sum_{i\in \cal U}\frac{1-2b_i^{\dag}b_i}{2\epsilon_i - E_{\nu}} 
\nonumber \\
&= E_{\nu}B_{\nu}^{\dag} 
+ \sum_{i\in \cal U}b_i^{\dag} 
- gB_0^{\dag}\sum_{i\in \cal U}\frac{1-2b_i^{\dag}b_i}{2\epsilon_i - E_{\nu}} 
\nonumber \\
&= E_{\nu}B_{\nu}^{\dag} + B_0^{\dag} - gB_0^{\dag}\sum_{i\in \cal U}\frac{1-2b_i^{\dag}b_i}{2\epsilon_i-E_{\nu}}
 = E_{\nu}B_{\nu}^{\dag} + B_0^{\dag}\left[1-g\sum_{i\in \cal U}\frac{1-2b_i^{\dag}b_i}{2\epsilon_i-E_{\nu}}\right]. 
\end{align}
Substituting this result into (\ref{eq: rich-hamiltonian_prod}) yields
\begin{align}\label{wyraz}
\sum_{\nu=1}^{M}&\left(\prod_{\eta=1}^{\nu-1}B_{\eta}^{\dag}\right)\left\{E_{\nu}B_{\nu}^{\dag}+B_0^{\dag}\left[1-g\sum_{j\in 
\cal U}\frac{1-2b_j^{\dag}b_j}{2\epsilon_j - E_{\nu}}\right]\right\}\left(\prod_{\mu=\nu+1}^{M}B_{\nu}^{\dag}\right)\ket{0} 
\nonumber\\
&= \sum_{\nu=1}^{M}\left(\prod_{\eta=1}^{\nu-1}B_{\eta}^{\dag}\right)E_{\nu}B_{\nu}^\dag\left(\prod_{\mu=\nu+1}^{M}B_{\mu}^{\dag}\right)\ket{0} + \sum_{\nu=1}^{M}\left(\prod_{\eta=1}^{\nu-1}B_{\eta}^{\dag}\right)B_0^{\dag}\left(\prod_{\mu=\nu+1}^{M}B_{\mu}^{\dag}\right)\ket{0} 
\nonumber\\
&-\sum_{\nu=1}^{M}\left(\prod_{\eta=1}^{\nu-1}B_{\eta}^{\dag}\right)B_0^{\dag}\sum_{j\in \cal U}\frac{g}
{2\epsilon_j - E_{\nu}}\left(\prod_{\mu=\nu+1}^{M}B_{\mu}^{\dag}\right)\ket{0} 
\nonumber\\
&+\sum_{\nu=1}^{M}\left(\prod_{\eta=1}^{\nu-1}B_{\eta}^{\dag}\right)\sum_{j\in \cal U}\frac{2gB_0^{\dag}b_j^{\dag}b_j}{2\epsilon_j - E_{\nu}}\left(\prod_{\mu=\nu+1}^{M}B_{\mu}^{\dag}\right)\ket{0}. 
\end{align}
Let us examine this expression term by term. The first term is
\begin{align}
\sum_{\nu=1}^{M}E_{\nu}&\left(\prod_{\eta=1}^{\nu-1}B_{\eta}^{\dag}\right)B_{\nu}^\dag\left(\prod_{\mu=\nu+1}^{M}B_{\mu}^{\dag}
\right)\ket{0} 
=\sum_{\nu=1}^{M}E_{\nu}\prod_{\nu=1}^{M}B_{\nu}^{\dag}\ket{0}
\nonumber\\
&= \sum_{\nu=1}^{M}E_{\nu}\ket{\Psi_{M}}_{\cal U}=\mathbb{E}(M)\ket{\Psi_{M}}_{\cal U}.
\end{align}
Since $[B_0^{\dagger},B_J^{\dagger}]=0$, we may pull $B_0^{\dagger}$ outside in the second term, which yields
\begin{equation}
\sum_{\nu=1}^{M}B_0^{\dag}\left(\prod_{\eta=1}^{M}B_{\eta}^{\dag}\right)\left(\prod_{\mu=\nu+1}^{M}B_{\mu}^{\dag}\right)\ket{0}
=\sum_{\nu=1}B_0^{\dag}\left(\prod_{\mu\neq \nu}^{M}B_{\mu}^{\dag}\right)\ket{0}.
\end{equation}
The third term reduces to
\begin{align}
\sum_{\nu=1}^{M}\left(\prod_{\eta=1}^{\nu-1}B_{\eta}^{\dag}\right)\sum_{j\in \cal U}\frac{g}{2\epsilon_j - E_{\nu}}
\left(\prod_{\mu=\nu+1}^{M}B_{\mu}^{\dag}\right)\ket{0} = \sum_{\nu=1}^{M}\sum_{j\in \cal U}\frac{g}{2\epsilon_j-E_{\nu}}
\left(\prod_{\mu\neq \nu}^{M} B_{\nu}^{\dag}\right)\ket{0},
\end{align}
which can then be combined with the second term to yield
\begin{equation}
\sum_{\nu=1}^{M}B_0^{\dag}\left[1-\sum_{j\in \cal U}\frac{g}{2\epsilon_j-E_{\nu}}\right]\left(\prod_{\mu\neq\nu}^{M}B_{\nu}
^{\dag}\right)\ket{0}.
\end{equation}
Finally, let us evaluate the last term by first considering the following commutator:
\begin{eqnarray}
\left[\sum_{j\in \cal U}\frac{2gB_0^{\dag}b_j^{\dag}b_j}{2\epsilon_j - E_{\nu}}, B_{\mu}^{\dag}\right] 
&=& 
\sum_{j\in \cal U}\frac{2gB_0^{\dag}b_j^{\dag}b_j}{2\epsilon_j - E_{\nu}}B_{\mu}^{\dag} - B_{\mu}^{\dag}\sum_{j\in \cal U}\frac{2gB_0^{\dag}b_j^{\dag}b_j}{2\epsilon_j - E_{\nu}}
\nonumber\\
&=& \sum_{j\in \cal U}\frac{\left(2gB_0^{\dag}b_j^{\dag}b_jB_{\mu}^{\dag}\right)- \left(2gB_{\mu}^{\dag}B_0^{\dag}b_j^{\dag}b_j\right)}{2\epsilon_j - E_{\mu}} 
\nonumber \\
&=&\sum_{j\in \cal U}\frac{\left(2gB_0^{\dag}b_j^{\dag}b_jB_{\mu}^{\dag}\right)- \left(2gB_0^{\dag}B_{\mu}^{\dag}b_j^{\dag}b_j\right)}{2\epsilon_j - E_{\mu}} 
\nonumber \\
&=&\sum_{j\in \cal U}\frac{2gB_0^{\dag}\left[b_j^{\dag}b_j, B_{\nu}^{\dag}\right]}{2\epsilon_j - E_{\nu}} 
\nonumber \\
&=& \sum_{j\in \cal U}\frac{2gB_0^{\dag}}{2\epsilon_j-E_{\nu}}\frac{b_j^{\dag}}{2\epsilon_i - E_{\mu}} 
\nonumber \\
&=& 2gB_0^{\dag}\sum_{j\in \cal U}\frac{b_j^{\dag}\left(E_{\nu}-E_{\mu}\right)}{\left(2\epsilon_j-E_{\nu}\right)\left(2\epsilon_{j}-E_{\mu}\right)\left(E_{\nu}-E_{\mu}\right)} 
\nonumber \\
&=& 2gB_0^{\dag}\sum_{j\in \cal U}\frac{b_{j}^{\dag}\left(2\epsilon_{j}-E_{\mu}\right) - b_{j}^{\dag}\left(2\epsilon_j - E_{\nu}\right)}{\left(2\epsilon_j-E_{\nu}\right)\left(2\epsilon_{j}-E_{\mu}\right)\left(E_{\nu}-E_{\mu}\right)} 
\nonumber \\
&=& 2gB_0^{\dag}\sum_{j\in \cal U}\left(\frac{b_j^{\dag}}{\left(2\epsilon_j-E_{\nu}\right)\left(E_{\nu}-E_{\mu}\right)} - \frac{b_j^{\dag}}{\left(2\epsilon_j-E_{\mu}\right)\left(E_{\nu}-E_{\mu}\right)}\right) 
\nonumber \\
&=& 2gB_0^{\dag}\frac{B_{\nu}^{\dag}-B_{\mu}^{\dag}}{E_{\nu}-E_{\mu}}.
\end{eqnarray}
This leads us to conclude that
\begin{align}
&\left[\sum_{j\in \cal U}\frac{2gB_0^{\dag}b_j^{\dag}b_j}{2\epsilon_j - E_{\nu}}, \prod_{\mu=\nu+1}^{M}B_{\mu}^{\dag}\right] = 
\sum_{\mu=\nu+1}^{M}\left(\prod_{\eta'=\nu+1}^{\mu-1}B_{J_{\eta'}}^{\dag}\right)\left[\sum_{j\in \cal U}\frac{2gB_0^{\dag}
b_j^{\dag}b_j}{2\epsilon_j - E_{\nu}}, B_{\mu}^{\dag}\right]
\nonumber\\
 &\times\left(\prod_{\mu'=\mu+1}^{M}B_{J_{\mu'}}^{\dag}\right) = \sum_{\mu=\nu+1}^{M}\left(\prod_{\eta'=\nu+1}^{\mu-1}
B_{J_{\eta'}}^{\dag}\right)2gB_0^{\dag}\frac{B_{\nu^{\dag}}-B_{\mu}^{\dag}}{E_{\nu}-E_{\mu}}
\left(\prod_{\mu'=\mu+1}^{M}B_{J_{\mu'}}^{\dag}\right)  
\nonumber\\
&=\sum_{\mu=\nu+1}^{M}\left(\prod_{\eta'=\nu+1}^{\mu-1}B_{J_{\eta'}}^{\dag}\right)\frac{2gB_0^{\dag}B_{\nu^{\dag}}}
{E_{\nu}-E_{\mu}}\left(\prod_{\mu'=\mu+1}^{M}B_{J_{\mu'}}^{\dag}\right) 
\nonumber\\
&-\sum_{\mu=\nu+1}^{M}\left(\prod_{\eta'=\nu+1}^{\mu-1}B_{J_{\eta'}}^{\dag}\right)
\frac{2gB_0^{\dag}B_{\mu^{\dag}}}{E_{\nu}-E_{\mu}}\left(\prod_{\mu'=\mu+1}^{M}B_{J_{\nu'}}^{\dag}\right) 
\nonumber\\
&=\sum_{\mu=\nu+1}^{M}\frac{2gB_0^{\dag}}{E_{\nu}-E_{\mu}}\left(\prod_{\eta'=\nu+1}^{\mu-1}B_{J_{\eta'}}^{\dag}
\right)B_{\nu}^{\dag}\left(\prod_{\mu'=\mu+1}^{M}B_{J_{\mu'}}^{\dag}\right)
\nonumber\\
&-\sum_{\mu=\nu+1}^{M}\frac{2gB_0^{\dag}}{E_{\nu}-E_{\mu}}
\left(\prod_{\eta'=\nu+1}^{\mu-1}B_{J_{\eta'}}^{\dag}\right)B_{\mu}^{\dag}
\left(\prod_{\mu'=\mu+1}^{M}B_{J_{\mu'}}^{\dag}\right).
\end{align}
Observe that the same commutator can also be expressed as
\begin{align}
\left[\sum_{j\in \cal U}\frac{2gB_0^{\dag}b_j^{\dag}b_j}{2\epsilon_j - E_{\nu}}, \prod_{\mu=\nu+1}^{M}B_{\mu}^{\dag}\right] 
&=\sum_{j\in \cal U}\frac{2gB_0^{\dag}b_j^{\dag}b_j}{2\epsilon_j - E_{\nu}}
\left(\prod_{\mu=\nu+1}^{M}B_{\mu}^{\dag}\right) 
\nonumber\\
&-\left(\prod_{\mu=\nu+1}^{M}B_{\mu}^{\dag}\right)\sum_{j\in \cal U}\frac{2gB_0^{\dag}b_j^{\dag}b_j}{2\epsilon_j - E_{\nu}} 
\;\;\;\;\Rightarrow 
\nonumber\\
&\hspace{-120pt}
\left[\sum_{j\in \cal U}\frac{2gB_0^{\dag}b_j^{\dag}b_j}{2\epsilon_j - E_{\nu}}, \prod_{\mu=\nu+1}^{M}B_{\mu}^{\dag}\right]\ket{0} = \sum_{j\in \cal U}\frac{2gB_0^{\dag}b_j^{\dag}b_j}{2\epsilon_j - E_{\nu}}
\left(\prod_{\mu=\nu+1}^{M}B_{\mu}^{\dag}\right)\ket{0}.
\end{align}
This means that our last term of (\ref{wyraz}) can be written as
\begin{align}
&\sum_{\nu=1}^{M}\left(\prod_{\eta=1}^{\nu-1}B_{\eta}^{\dag}\right)\left[\sum_{j\in \cal U}\frac{2gB_0^{\dag}b_j^{\dag}b_j}
{2\epsilon_j - E_{\nu}}, \prod_{\mu=\nu+1}^{M}B_{\mu}^{\dag}\right]\ket{0} 
\nonumber\\
 &=\sum_{\nu=1}^{M}\left[\sum_{\mu=\nu+1}^{M}\frac{2gB_0^{\dag}}{E_{\nu}-E_{\mu}}\left(\prod_{\eta=1}^{\nu-1}B_{\eta}^{\dag}\right)\left(\prod_{\eta'=\nu+1}^{\mu-1}B_{J_{\eta'}}^{\dag}\right)B_{\nu}^{\dag}\left(\prod_{\mu'=\mu+1}B_{J_{\mu'}}^{\dag}\right)\right]\ket{0} 
\nonumber\\
&-\sum_{\nu=1}^{M}\left[\sum_{\mu=\nu+1}^{M}\frac{2gB_0^{\dag}}{E_{\nu}-E_{\mu}}\left(\prod_{\eta=1}^{\nu-1}B_{\eta}^{\dag}\right)\left(\prod_{\eta'=\nu+1}^{\mu-1}B_{J_{\eta'}}^{\dag}\right)B_{\mu}^{\dag}\left(\prod_{\mu'=\mu+1}B_{J_{\mu'}}^{\dag}\right)\right]\ket{0} 
\nonumber\\
&=\sum_{\nu=1}^{M}\left[\sum_{\mu=\nu+1}^{M}\frac{2gB_0^{\dag}}{E_{\nu}-E_{\mu}}\left(\prod_{\eta\neq \mu}^{M}B_{\eta}^{\dag}\right)\right]\ket{0} - \sum_{\nu=1}^{M}\left[\sum_{\mu=\nu+1}^{M}\frac{2gB_0^{\dag}}{E_{\nu}-E_{\mu}}\left(\prod_{\eta\neq \nu}^{M}B_{\eta}^{\dag}\right)\right]\ket{0} 
\nonumber\\
&=\sum_{\mu=1}^{M}\left[\sum_{\nu=1}^{\mu-1}\frac{2gB_0^{\dag}}{E_{\nu}-E_{\mu}}\right]\left(\prod_{\eta\neq \mu}^{M}B_{\eta}^{\dag}\right)\ket{0} - \sum_{\nu=1}^{M}\left[\sum_{\mu=\nu+1}^{M}\frac{2gB_0^{\dag}}{E_{\nu}-E_{\mu}}\right]\left(\prod_{\eta\neq \nu}^{M}B_{\eta}^{\dag}\right)\ket{0} 
\nonumber \\
&=\sum_{\nu=1}^{M}\left[\sum_{\mu=1}^{\nu-1}\frac{2gB_0^{\dag}}{E_{\mu}-E_{\nu}}\right]\left(\prod_{\eta\neq \nu}^{M}B_{\eta}^{\dag}\right)\ket{0} - \sum_{\nu=1}^{M}\left[\sum_{\mu=\nu+1}^{M}\frac{2gB_0^{\dag}}{E_{\nu}-E_{\mu}}\right]\left(\prod_{\eta\neq \nu}^{M}B_{\eta}^{\dag}\right)\ket{0} 
\nonumber\\
&=\sum_{\nu=1}^{M}\left[\sum_{\mu-1}^{\nu-1}\frac{2gB_0^{\dag}}{E_{\mu}-E_{\nu}} + \sum_{\mu=\nu+1}^{M}\frac{2gB_0^{\dag}}{E_{\mu}-E_{\nu}}\right]\left(\prod_{\eta\neq \nu}^{M}B_{\eta}^{\dag}\right)\ket{0} 
\nonumber\\
&=\sum_{\nu=1}^{M}B_0^{\dag}\left[\sum_{\mu\neq \nu}\frac{2g}{E_{\mu}-E_{\nu}}\right]
\left(\prod_{\eta\neq \nu}^{M}B_{\eta}^{\dag}\right)\ket{0}. 
\end{align}
Collecting all terms, we obtain
\begin{align}
H_{\cal U}\ket{\Psi_M}_{\cal U} 
&= \mathbb{E}(M)\ket{\Psi_M}_{\cal U} + \sum_{\nu=1}^{M}B_0^{\dag}\left[1-\sum_{j\in \cal U}\frac{g}
{2\epsilon_j-E_{\nu}}\right]\left(\prod_{\eta\neq \nu}^{M}B_{\eta}^{\dag}\right)\ket{0} 
\nonumber\\
&\hspace{80pt}
+\sum_{\nu=1}^{M}B_0^{\dag}\left[\sum_{\mu\neq \nu}\frac{2g}{E_{\mu}-E_{\nu}}\right]\left(\prod_{\eta\neq \nu}
^{M}B_{\eta}^{\dag}\right)\ket{0} 
\nonumber\\
&\hspace{-50pt}
=\mathbb{E}(M)\ket{\Psi_M}_{\cal U} + \sum_{\nu=1}^{M}B_0^{\dag}\left[1-\sum_{j\in \cal U}\frac{g}{2\epsilon_j-E_{\nu}} + 
\sum_{\mu\neq \nu}\frac{2g}{E_{\mu}-E_{\nu}}\right]\left(\prod_{\eta\neq \nu}^{M}B_{\eta}^{\dag}\right)\ket{0}. 
\end{align}

\section{Rank--one irregular state and irregular vertex operator}
\label{Irr1}
Let $r$ denote the rank---the largest integer for which the eigenvalue of the Virasoro generator 
$L_{2r}$ on the irregular state is nonzero. In the rank-one case ($r=1$), the irregular state 
$\ket{I_{\alpha,\gamma}}$ satisfies the following conditions \cite{G13,GT12}:
\begin{align}
L_{0}\ket{I_{\alpha\gamma}}
&=\bigl(\Delta_{\alpha}+\gamma\partial_{\gamma}\bigr)\ket{I_{\alpha,\gamma}},
\label{L0}\\
L_{1}\ket{I_{\alpha,\gamma}}
&=\gamma\bigl(Q-\alpha\bigr)\ket{I_{\alpha,\gamma}},\\
L_{2}\ket{I_{\alpha,\gamma}}
&=-\tfrac{\gamma^{2}}{4}\ket{I_{\alpha,\gamma}},
\label{L2}
\end{align}
where
$$
\Delta_{\alpha}=\alpha\,(Q-\alpha), 
\qquad
Q=b+\frac{1}{b},
$$
and $\gamma$ is an arbitrary complex parameter.

The irregular state $\ket{I_{\alpha,\gamma}}$
meeting (\ref{L0})-(\ref{L2}) is generated from the vacuum $\ket{0}$ by the 
corresponding rank-one irregular chiral vertex operator (CVO)~\cite{HLR},
\begin{equation}\label{I}
I_{\alpha,\gamma}(w)\equiv:{\rm e}^{2\alpha\varphi(w)}{\rm e}^{\gamma\partial\varphi(w)}:,
\end{equation}
according to the operator-state correspondence, i.e.,
\begin{equation}\label{PSC}
\lim\limits_{w\to 0}I_{\alpha,\gamma}(w)\ket{0}=\ket{I_{\alpha,\gamma}}.
\end{equation}
The rank-one CVO (\ref{I}) consists of ordered exponentials, where $\varphi(w)$ is the free field
obeying
\begin{equation}\label{Propa}
\langle\varphi(z)\varphi(w)\rangle=-\frac{1}{2}\log(z-w).
\end{equation}
In order to validate that the operator $I_{\alpha,\gamma}(w)$ generates the 
irregular state $\ket{I_{\alpha,\gamma}}$ it is necessary to compute 
its OPE with the holomorphic component of the energy-momentum tensor,\footnote{The form (\ref{Tz}) of $T(z)$ 
corresponds to 2D CFT with the central charge $c=1+6Q^2$.}
\begin{equation}\label{Tz}
T(z)=-:\partial\varphi(w)\partial\varphi(w):+Q\partial^{2}\varphi(w),
\end{equation}
and calculate the commutation relations of $I_{\alpha,\gamma}(w)$ with $L_0$, $L_1$ and $L_2$.
Recall that commutators of the Virasoro generators $L_n$ with the vertex operators $V(w)$ are given by the contour integral,
\begin{eqnarray}\label{com}
\left[L_n,V(w)\right]&=&
\oint_{C_w}\frac{{\rm d}z}{2\pi i}z^{n+1}T(z)V(w),
\end{eqnarray}
where the contour $C_w$ surrounds $w$. Indeed, from (\ref{Propa}) and Wick's 
theorem one can find the leading singular behavior of the aforementioned OPE:
\begin{eqnarray}\label{TI}
T(z)I_{\alpha,\gamma}(w)&=&
\left(\frac{2\alpha}{z-w}\partial\varphi(w)+
\frac{\gamma}{(z-w)^2}\partial\varphi(w)+
\frac{\alpha(Q-\alpha)}{(z-w)^2}\right.\nonumber\\
&+&\left.
\frac{\gamma(Q-\alpha)}{(z-w)^3}
-\frac{\gamma^2}{4}\frac{1}{(z-w)^4}
\right)I_{\alpha,\gamma}(w).
\end{eqnarray}
Next, by making use of the OPE (\ref{TI}) and shifting the integration variable $z\to z+w$ in (\ref{com}) 
one can compute the commutators:
\begin{itemize}
\item[---] for $n=0$
\begin{eqnarray}\label{n0}
\left[L_0,I_{\alpha,\gamma}(w)\right]&=&
\oint_{C_0}\frac{{\rm d}z}{2\pi i}\left(z+w\right)
\left(\frac{2\alpha}{z}\partial\varphi(w)+
\frac{\gamma}{z^2}\partial\varphi(w)+
\frac{\Delta_\alpha}{z^2}\right.\nonumber\\
&+&\left.
\frac{\gamma(Q-\alpha)}{z^3}
-\frac{\gamma^2}{4}\frac{1}{z^4}
\right)
I_{\alpha,\gamma}(w)\nonumber\\
&=& \left(2\alpha w\partial\varphi(w)+\gamma\partial_{\gamma}+\Delta_{\alpha}
\right)I_{\alpha,\gamma}(w);
\end{eqnarray}
\item[---] for $n=1$
\begin{eqnarray}\label{n1}
\left[L_1,I_{\alpha,\gamma}(w)\right]&=&
\oint_{C_0}\frac{{\rm d}z}{2\pi i}\left(z+w\right)^2
\left(\frac{2\alpha}{z}\partial\varphi(w)+
\frac{\gamma}{z^2}\partial\varphi(w)+
\frac{\Delta_\alpha}{z^2}\right.\nonumber\\
&+&\left.
\frac{\gamma(Q-\alpha)}{z^3}
-\frac{\gamma^2}{4}\frac{1}{z^4}
\right)
I_{\alpha,\gamma}(w)\nonumber\\
&&\;\hspace{-50pt}=\left(2\alpha w^2\partial\varphi(w)+2\gamma w\partial\varphi(w)
+2w\Delta_{\alpha}+\gamma(Q-\alpha)
\right)I_{\alpha,\gamma}(w);
\end{eqnarray}
\item[---] for $n=2$
\begin{eqnarray}\label{n2}
\left[L_2,I_{\alpha,\gamma}(w)\right]&=&
\oint_{C_0}\frac{{\rm d}z}{2\pi i}\left(z+w\right)^3
\left(\frac{2\alpha}{z}\partial\varphi(w)+
\frac{\gamma}{z^2}\partial\varphi(w)+
\frac{\Delta_\alpha}{z^2}\right.\nonumber\\
&+&\left.
\frac{\gamma(Q-\alpha)}{z^3}
-\frac{\gamma^2}{4}\frac{1}{z^4}
\right)
I_{\alpha,\gamma}(w)\nonumber\\
&&\;\hspace{-80pt}=\left(2\alpha w^3\partial\varphi(w)+3\gamma w^2\partial\varphi(w)
+3w^2\Delta_{\alpha}+3w\gamma(Q-\alpha)-\frac{\gamma^2}{4}
\right)I_{\alpha,\gamma}(w).
\end{eqnarray}
\end{itemize}
Finally, by taking the limit 
$$
\lim_{w\to 0}L_nI_{\alpha,\gamma}(w)\ket{0}
=\lim_{w\to 0}\left[L_n,I_{\alpha,\gamma}(w)\right]\ket{0}
$$ 
for $n=1,2,3$  and implementing (\ref{n0})-(\ref{n2}) alongside (\ref{PSC}), the conditions (\ref{L0})-(\ref{L2}) are reproduced.

\end{document}